\documentclass[twocolumn,times,tighten]{aastex631}

\usepackage{natbib}
\usepackage{graphicx}

\usepackage{ulem}

\newcommand{\del}[1]{\textcolor{magenta}{{\iffalse{#1}\fi}}}

\begin{document} 

\title{Pulsar B1237+25 at 111~MHz: average profile, mode switching, nullings, microstructure}
   
\author{M.V. Popov}
\affiliation{Astrospace Center, Lebedev Physical Institute, Astro Space Center, Russian Academy of Science, Moscow,117997 Russia}
\author{T.V. Smirnova}
\affiliation{Lebedev Physical Institute, Astro Space Center, Pushchino Radio Astronomy Observatory, 
142290, Moscow region, Pushchino, Russia}

\date{Received ; accepted}
   
\begin{abstract}
The observations of pulsar B1237+25 at a frequency of 111 MHz are analyzed.
 The observations were carried out with the Large Phase Array (LPA) of Pushchino Radio Astronomical Observatory, Lebedev Physical Institute. For the first time in the normal radiation mode a new component in the central region of the emission was detected in the average profile. This component with a half-width of 1.6~ms is observed against the background of the usual longer component with a half-width of 7~ms. This new component of the average profile at a frequency of 111~MHz is manifested in all modes of pulsar radio emission:
 quiet-normal (QN), flare-normal (FN) and in the abnormal mode (AB), while
 in the abnormal mode it was detected by other observers at other frequencies.

The subpulse drift is observed in the normal QN  mode only in the first and last components of the average profile. The normal mode is interrupted by nullings and transitions into
the abnormal AB mode. In the AB mode, the structure of the activity zones at the edge of the outer cone is destroyed, the distance between the inner and outer cones is almost doubled, and the distance between the inner cone and the central region is reduced.

 Analysis of our data has shown that the components of the outer and inner cones of the average profile are formed by an ordinary mode of radio emission (O-mode) and form a single cone radiation of the pulsar. The central components of the average profile (wide and narrow) are formed by an extraordinary radiation mode (X-mode). Estimates of the height of the radiation output from the central region (X-mode) and the cone radiation (O-mode) are obtained: 80~km and
370~km, respectively.
 
A microstructure with a submicrosecond time scale of $\tau_\mu\le0.5$~microseconds has been detected. This time scale corresponds well to the characteristic time of the development of a spark discharge in the polar cap. For the value we have defined
$\tau_\mu$ the height of the vacuum gap should be $h_p\le750$~cm. Based on the steepness of the individual pulse's trailing edge at the longitude of the first component, a limit was obtained on the value of the $\gamma$ factor of the relativistic secondary plasma: $\gamma\ge$260.

The analysis showed that in 88\% of cases the normal mode (N) is realized, and in 81\% of them the QN mode and only in 19\% the FN mode. The AB mode is only 12\%. The dependence of the distance between the components of the outer and inner cone of radiation on the frequency is the same and corresponds to a power law with an exponent of -0.16.

\end{abstract}

\section{Introduction} 

 The pulsar B1237+25 has been studied in many works since the 70s: \citep{backer1970}, \citep{rankin1986}, \citep{srost2005}, \citep{smith2013}. 
   According to the Rankin classification~\citep{rankin1993} the average profile consists of five components,
   which reflect the presence of
   two concentric radiation cones and a central region (‘core beam'). 
   According to the data presented in the work of Smith and co-authors \citep{smith2013}, the angle $\zeta$ between the axis of rotation and the magnetic axis is $57.6^{\circ}$, and the line of sight passes almost through the center of two concentric cones and the core beam; the deviation of 
   the line of sight from the magnetic axis in the center of the profile is only $\beta{=}-0.3^{\circ}$. The pulsar is characterized by the phenomenon of
 mode switching when the profile shape changes significantly from "normal" (N) to "abnormal" (AB) mode\citep{backer1970}.  In the N mode, the 2 extreme components are clearly visible, and the central ones are weak. In this mode, a subpulse modulation with a period of 2.7 of the pulsar period $P_1$\citep{bartel1982} is observed.  Srostlik and Rankin~\citep{srost2005} found two different modes in the normal mode: the quiet normal mode (QN) with a regular $2.7P_1$ modullation in the leading and trailing components and with low activity in the center, and the flare-normal mode (FN), in which the central component is bright, but less than the 1st component, it has about 4 $P_1$ modulation. In the AB mode, there is a strong central component and significantly weaker side components. In addition, there is no periodic subpulse modulation.
 There is a normal flash mode (FN - flare-normal) with modulation in the extreme components and low activity in the center, in which the central component is bright, but less than the 1st component, it has about 4$P_1$ modulation. 
 
\citep{backer1973} constructed fluctuation spectra for 13 pulsars based on the results of observations at the Arecibo radio telescope at frequencies of 318, 430 and 606~MHz. He showed that isolated maxima in the fluctuation spectra are associated with the phenomenon of regular drift of subpulses along longitude within the average profile. Such a drift is usually characterized by additional parameters: the period $P_2$ indicates the distance in longitude
between the drifting subpulses, and the period $P_3$ corresponds to the time of repetition of the drift pattern.
 In the fundamental work of Ruderman and Sutherland~\citep{1975ApJ...196...51R}
 an explanation of the phenomenon of subpulse drift was presented as a result of the displacement of the discharge in the vacuum gap of the polar cap, which generates a secondary plasma responsible for radio emission. The process of avalanche-like generation of secondary plasma was first considered by Sturrock~\citep{1971ApJ...164..529S}. An important feature in the formation of electric discharges in the vacuum gap of the polar cap was noted by
 Beskin~\citep{1982SvA....26..443B}; he showed that the adjacent discharges must be located at a certain distance due to mutual screening by their own electric field. Thus, there may be a limited number of discharges in the polar cap, about a dozen.

 The phenomenon of subpulse drift has been intensively studied for more than 50 years after the discovery of pulsars. A fairly detailed review of the results of such studies and a discussion of the problem are presented by Rankin~\citep{1986ApJ...301..901R}. She also introduced the concept of a "carousel" 
 drifting subpulses using the example of pulsar B0943+10~\citep{1999ApJ...524.1008D}. In this model, discharges located
on a regular basis at the boundary of the radiation cone, rotate around a magnetic axis in crossed electric and magnetic fields. If the section
of the cone passes along its boundary with the line of sight, then several subpulses separated by a time interval $P_2$ in longitude can be observed simultaneously.  
 In the case of the central section of the cone of radiation crossed by the line of sight, the subpulse drift occurs in a perpendicular direction, and the displacement of  subpulses in longitude is not observed. Only the intensity modulation is noticeable at the edge longitudes of the average profile with a characteristic period of $P_3$. This type of drift  is characteristic of the studied pulsar B1237+25.

 Maan and Deshpande~\citep{2014ApJ...792..130M} performed a detailed analysis
of the drift behavior of pulsar B1237+25 at a frequency of 327~MHz based on the results of observations at the Arecibo radio telescope. They estimated 
 the number of drifting active spots (10-20) and the rotation period of the entire carousel is $P_4$($20-30~P_1$). In the studies of Srostlic and Rankin ~\citep{2005MNRAS.362.1121S}
and Smith et al. ~\citep{2013MNRAS.435.1984S}, it was noted that the feature at a frequency of 0.37 cycles per period in the fluctuation spectrum is clearly manifested only in the quiet normal mode of QN radiation. Frequency 
 $0.37s/P_1$ corresponds to the value of $P_3{=}2.7P_1$. Kramer et al. \citep{Kramer2006} claim that mode switching is associated with global changes in the pulsar's magnetosphere.
  
 Polarization measurements (Srostlik and Rankin~\cite{srost2005}) show that the  behavior of the position angle in the center of the average pulsar profile has a sharp jump, which indicates the presence of radiation in two orthogonal polarization modes (OPM). Depolarization of radiation at the edges of the average profile is also due to the presence of radiation in two ortogonal modes (OPM) (Rankin, Ramachandran\citep{Rankin2003}). This leads to the conclusion that one of the orthogonal modes is shifted relative to the other in magnetic latitude and azimuth. It is assumed that the radiation directed parallel and perpendicular to the projection of the magnetic field is due to their different refractive properties.  Analysis of the pulsar's polarization measurements at a frequency of 327 MHz made it possible to determine the radiation heights of the outer and inner cones: 340 km and 278 km, respectively, as well as 60 km for the central region \citep{srost2005}.

In our study, we performed observations and analysis of the behavior of pulsed radiation in all three modes at a frequency of 111~MHz. The longitudinal distribution of pulse amplitudes, nullings, time variations of pulse amplitudes at different longitudes, as well as the frequency dependence of widths and distances between the components of the average pulsar profile were considered.\\

\section{Observations}\label{observe}

PSR B1237+25 is a nearby pulsar. The distance to it is 840 pc, the measure of dispersion is $DM {=}9.25~pc/cm^3$, the measure of rotation is $RM = 0.3~rad/m^2$ (ATNF Pulsar Catalog). Observations of B1237+25 were carried out with the Large Phased Array (LPA) Pushchino Radio Astronomical Observatory of the Russian Academy of Sciences at 111 MHz as part of the pulsar diffractive parameters monitoring program. The telescope receives linearly polarized radiation, and the receiver has a 2.5 MHz bandwidth. The time of one session corresponds to the time the source passes through the ½ of the diagram  and 
 is 210 seconds (152 pulses).  At the time corresponding to the arrival time of the 1st pulse, buffering of the data coming from the ADC begins. Blocks of 2048 samples are written to the hard disk continuously for a specified time. The time interval between samples (8-bit signed integers) is 0.2 microseconds. Then coherent dispersion compensation is performed. The work of Girin and co-authors~\citep{girin2023} provides a detailed description of this
 method. From the continuous recording, we selected windows with a duration of 160 ms, separated by the pulsar period. These windows included both the pulse itself and the noise, which were used to draw a baseline and determine the RMS deviation of the noise from the mean value $\sigma_N$. The value of $\sigma_N$ was determined outside the pulse window in an interval equal to the pulse window. 

We also used another type of recording, which does not require a high time resolution and, consequently, significantly faster data processing. The receiver band was divided into 512 channels and the data were recorded after the dispersion was removed for all channels in a window synchronized with the pulsar period. The window size and time resolution are selected as parameters in this mode. The maximum time resolution in this mode is 0.2 ms.

\begin{table}[!h]
 \caption{The parameters of the average profile in different modes.} 
    \label{tab:width}
    \centering
    \begin{tabular}{c*{7}{c}}
        \hline
Mode&&\multicolumn{4}{c}{Component number}&&\\
   \hline
   &1&2&3&4&5&6\\
   \hline
   &&\multicolumn{3}{c}{Halfwidth (ms)}&&\\
  Mean &3.18(4)&4.7(3)&7.0(1)&1.6(1)&3.3(1)&4.34(4)\\
   QN&2.56(1)& 6.9(1)& 5.7(2)&2.0(1)&3.4(1)&3.82(2)\\
 FN&3.34(2)&6.6(2)&4.7(1)&2.52(3)&3.1(2)&4.81(4)\\
 AB&4.06(5)&5.4(2)&3.33(5)&1.52(1)&&7.1(3)\\
\hline
 &&\multicolumn{3}{c}{Location (ms)}&&\\
  Mean
  &-0.04(3)&10.5(2)&30.67(8)&32.56(5)&44.6(1)&55.98(5)\\
   QN&-0.19(1)& 8.1(1)& 28.6(1)&32.9(1)&44.2(1)&55.7(2)\\
 FN&-0.68(2)&9.9(1)&28.47(8)&32.39(3)&44.2(1)&55.59(4)\\
 AB&-0.38(4)&16.1(2)&27.53(4)&31.88(1)&&56.8(2)\\
\hline
 &&\multicolumn{3}{c}{Amplitude (Jy)}&&\\
  Mean &6.48(5)&0.98(4)&0.72(1)&0.51(2)&0.79(3)&2.46(2)\\
  &1.0&0.15&0.11&0.08&0.12&0.38\\
   QN&11.4(4)&1.58(1)&0.81(2)&0.94(3)&1.15(2)&4.08(2)\\
   &1.0&0.14&0.07&0.08&0.10&0.36\\
 FN&5.05(3)&0.70(1)&1.24(2)&2.00(3)&0.47(2)&2.12(1)\\
 &1.0&0.14&0.25&0.40&0.09&0.42\\
 AB&3.65(3)&1.29(3)&4.14(4)&10.42(6)&&1.03(2)\\
 &1.0&0.35&1.14&2.85&&0.28\\
\hline
        \end{tabular}
        \end{table} 
Note that diffractive scintillation does not affect the shape of the average profile in session, since the decorrelation bandwidth $f_{dif}$  and the scintillation time $t_{dif}$ are significantly less than the receiver bandwidth (2.5~MHz) and the observation time (3.5 min). The parameters measured at a frequency of 102.5 MHz are: $f_{dif}$=5~kHz, $t_{dif}$=32~sec ~\cite{smir1992}. The period of the Faraday rotation of the plane of polarization during the propagation of radiation in the interstellar plasma at a frequency of 111 MHz is 79.6 MHz, which significantly exceeds the receiver band. This can lead to variations in the position of the pulsar profile from session to session.\\
To estimate the flux density of the detected radiation, we used the relative increment of the detected signal level, taking the system equivalent flux density (SEFD) of the radio telescope equal to 50~Jy for the zenith
angle of $30^{\circ}$ and the galactic latitude of the pulsar $86^{\circ}$.

\section{Data analysis and results}   
\subsection{Mean profile}

To obtain the average profile of B1237+25, we averaged the profiles of 69 sessions (10902 pulses) of observations. To eliminate their possible shifts in different sessions, a cross-correlation of the reference profile with all subsequent profiles was performed.
The average profile related to the normal mode and received in a session with a good signal-to-noise ratio was used as a reference profile. The individual profiles were shifted in time by an appropriate interval to obtain the maximum correlation coefficient with the reference.
  Then, the average profile was approximated by a set of Gaussian components using the least squares method.  The accumulated average profile with a resolution of 0.2 ms is shown in Fig.1 in the upper panel. For the zero longitude, we took the longitude at which the first component of the total profile has a maximum. 

At our frequency of 111~MHz, 2 components (wide and narrow) were found at the central longitudes. They offset by 1.88 ms in longitude, and they overlapped with each other during averaging. The positions, amplitudes and widths were determined at a level of 0.5 from the maximum for the inscribed components. Table 1 gives the amplitudes of the components in three modes (QN, FN, AB), their positions in longitude, and half-width (W0.5) in ms. The longitude was calculated relative to the maximum position of the first component. The numbers in parentheses indicate errors of the last sign of the corresponding values, determined by fitting. The amplitudes of the components are given in Jy (first row) and in relative units relative to the 1st component (second row). A comparison of these parameters with published measurements at other frequencies will be made in Section 4.
\begin{figure}
 \centering
 \includegraphics[angle=270, width=80mm]{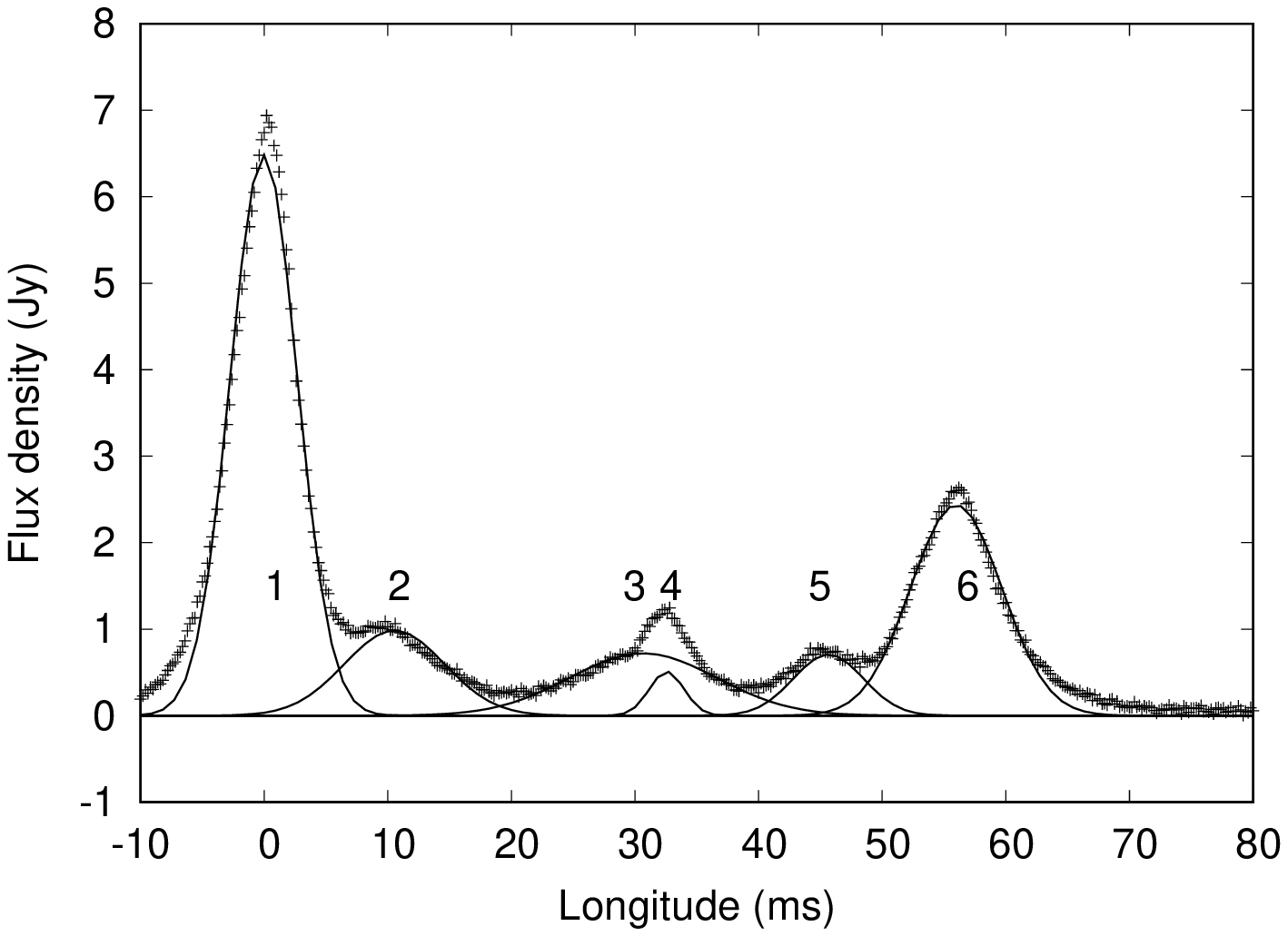}
\includegraphics[angle=270, width=80mm]{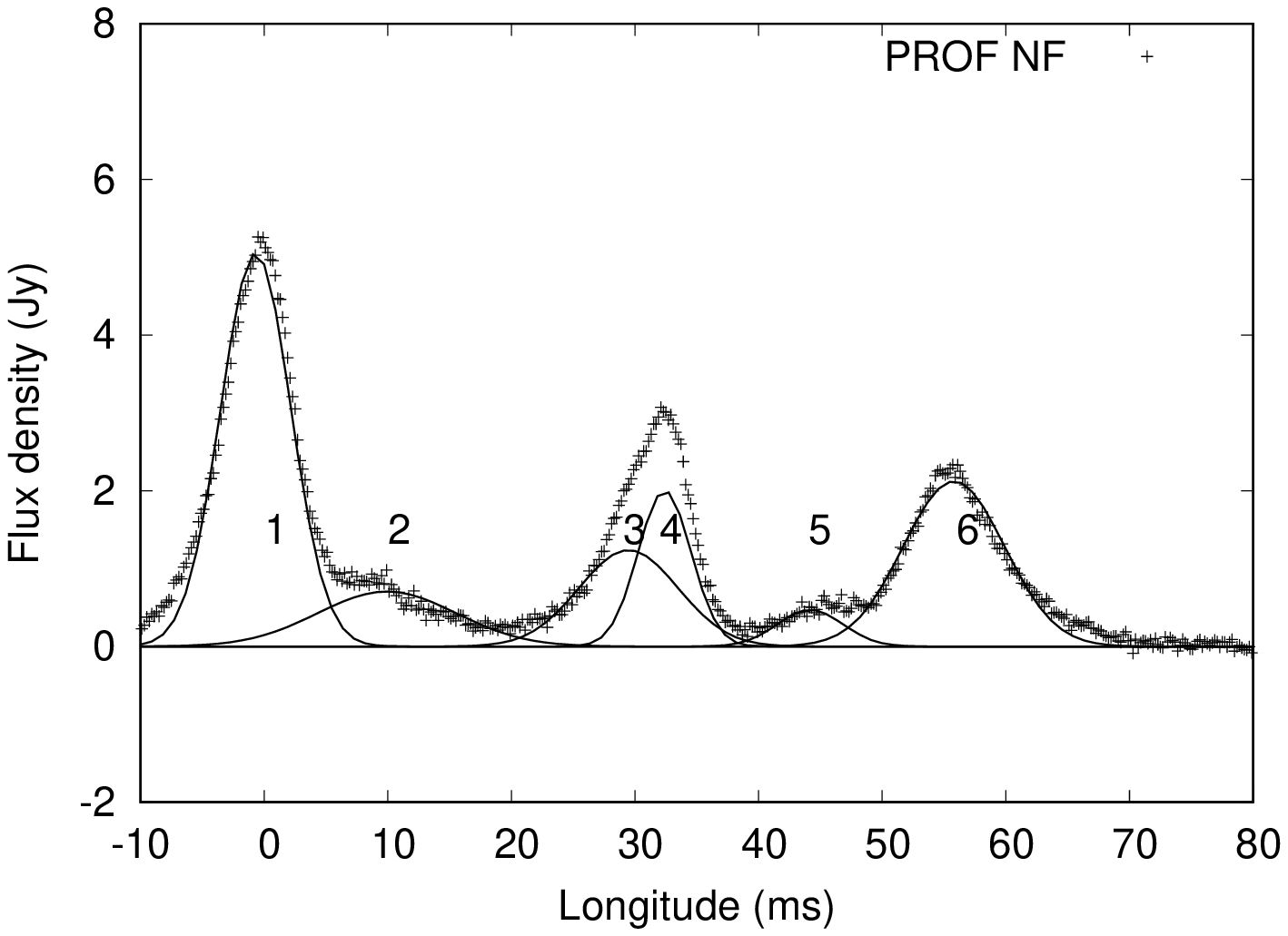}
\includegraphics[angle=270, width=80mm]{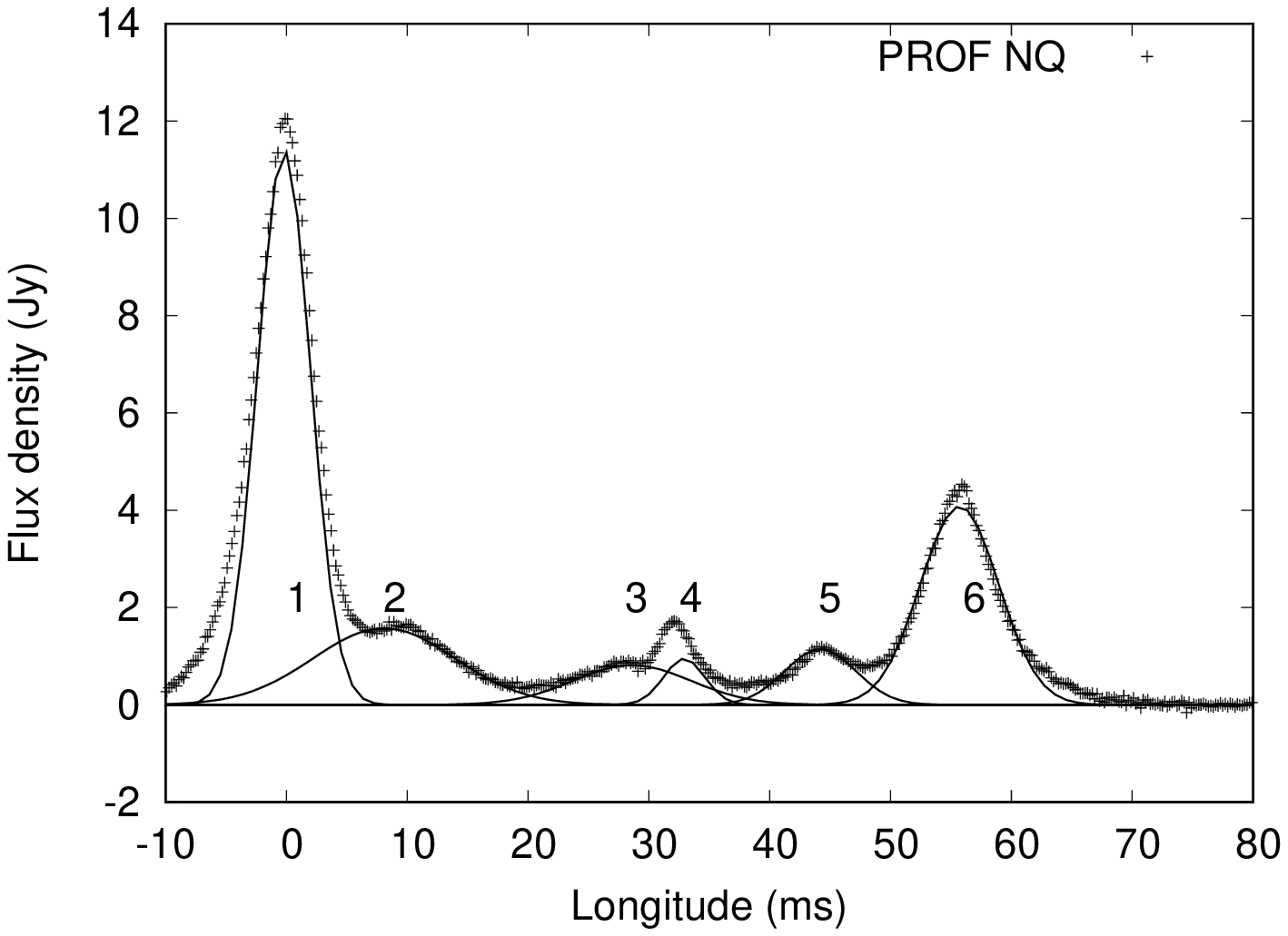}
\includegraphics[angle=270, width=80mm]{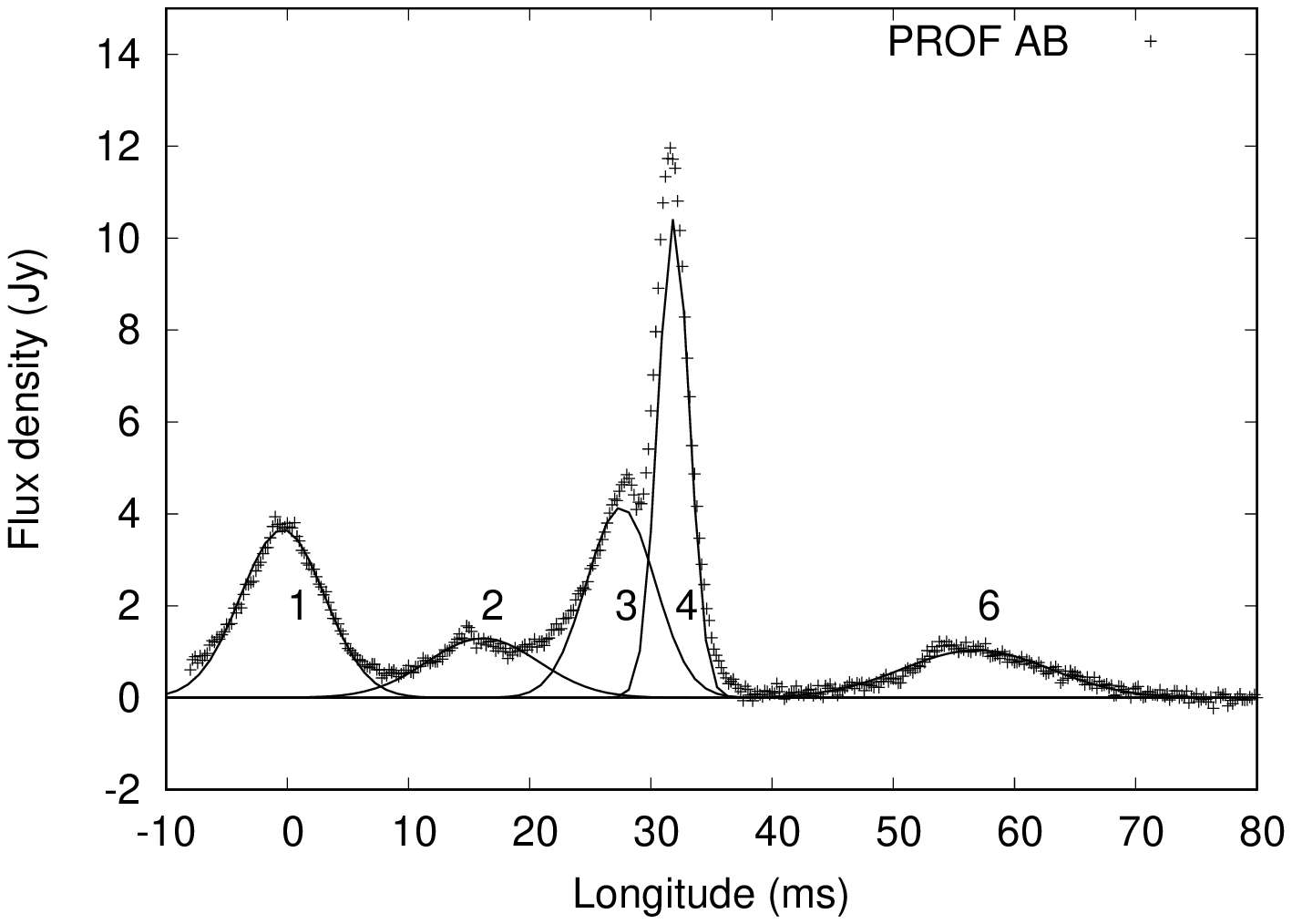}
 \caption{The average profile for all pulses (top figure), the number of averaged pulses N=10902; the average profile for the normal flare mode NF: N=2432 (second figure); the average profile for the normal quiet mode NQ: N=3040 (third),  and  profile for the anomalous mode AB: N=912 pulses (bottom)}
    \label{fig:fig1}%
    \end{figure}

Figure 1 also shows the average profiles constructed for different modes of pulsar radiation:
The normal quiet mode (QN) is on the third  picture, the normal flash mode (FN) is on the second one, and the average profile for the abnormal mode (AB) is shown on the bottom picture. When constructing average profiles in different radiation modes, only observational sessions were used in which the pulsar showed sufficient brightness for reliable classification of the radiation mode. In this case the average profile for the best session related to the corresponding radiation mode was used as a reference for the alignment of individual profiles in longitude.
As a result, the average profile for the QN mode was obtained using 3040 pulses, and the average profiles for the FN and AB modes contain 2432 and 912 pulses, respectively. The parameters of these average profiles are given in Table 1. 

An analysis of the profiles accumulated during the sessions showed that in 88\% of cases the normal mode (N) is implemented, and in 81\% of them being the QN mode and only 19\% being the FN mode. AB mode is only 12\%. 

 \subsection{Longitude distribution of pulse amplitudes}
 
Fig. 2 and 3 show the profiles for 2 observation sessions in the QN and FN modes with a time resolution of 1.43 ms. Figure 4 shows a session in the AB mode. The same figures show the positions of the maximum amplitudes of individual pulses exceeding the level of 3$\sigma_N$. In the normal quiet mode (QN), strong pulses are realized in the extreme components, while they are weak in the other components. In the FN mode, the strongest pulses fall into the first component. This mode is characterized by approximately the same distribution of amplitudes and the number of pulses at the longitudes of the central and last components (Fig. 3).

\begin{figure}
   \centering
   \includegraphics[width=0.8\columnwidth]{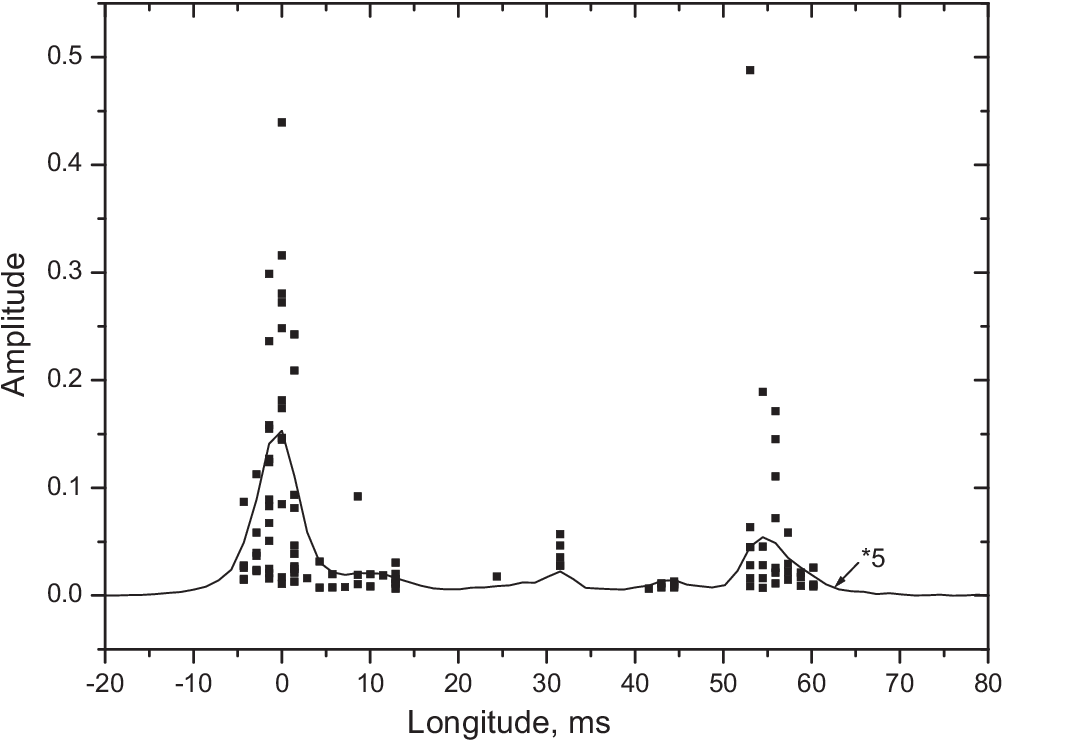}
   \caption{Figure 2. The total profile for 25.02.24 in the normal mode (QN), increased by 5 times, S/N = 371. The squares show the distribution of the maxima of individual pulses along the longitude.  x–axis is the longitude in ms, y–axis is the amplitude in relative units.
   }
    \label{fig:fig2n}%
    \end{figure}

\begin{figure}
   \centering
   \includegraphics[width=0.8\columnwidth]{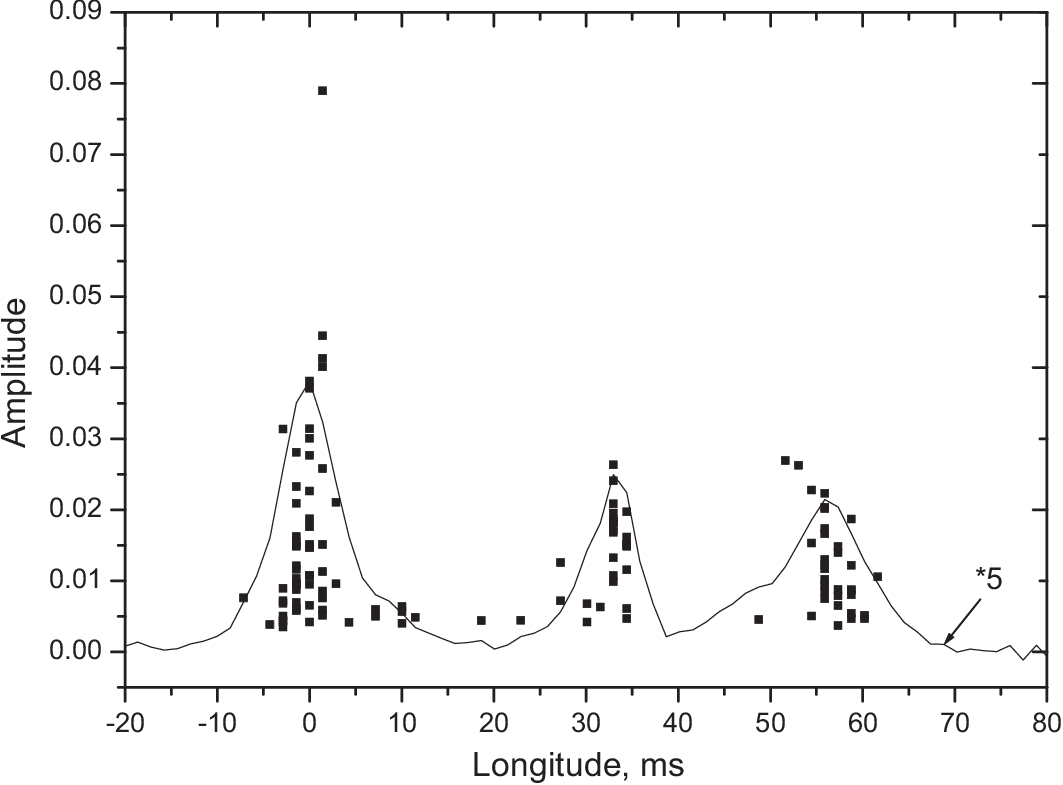}
   \caption{The total profile for 02.03.24 in the normal mode (FN), increased by 5 times, S/N = 66.2. The squares show the distribution of individual pulse maxima over longitude. The designation along the axes is the same as in Fig. 2.
   }
    \label{fig:fig3n}%
    \end{figure}    
To see how the profile changes with the accumulation of pulses in a limited range of amplitudes, we introduced an amplitude limit: a1*$\sigma_N$ and a2*$\sigma_N$, where a1 and a2 are the boundary amplitudes from below and above. Figure 5 shows the accumulated pulse profiles for the 25.02.24 (QN) mode session. In the normal mode in this session, when obtaining a profile with pulse amplitudes in the range of 3 — 30$\sigma_N$, radiation is present at the longitudes of all components and they have approximately equal amplitudes in the profile. The amplitude of this profile is
\begin{figure}
\centering
\includegraphics[angle=0, width=80mm]{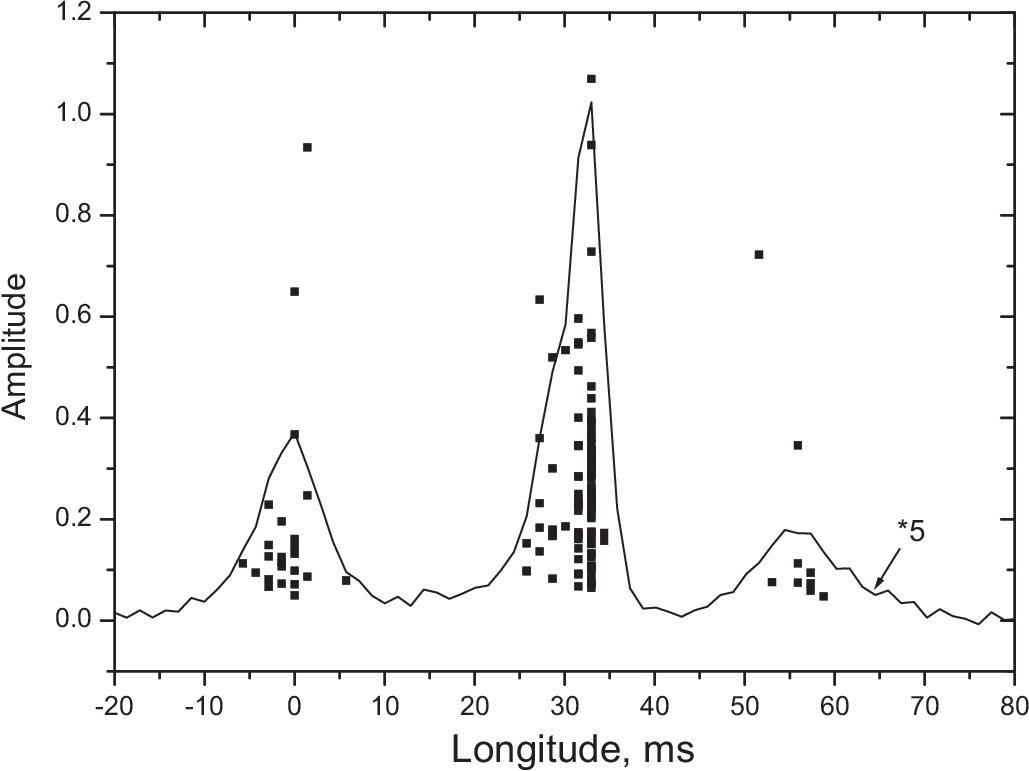}
\caption{The total profile for 01.01.24 in the abnormal mode (AB), increased by 5 times, S/N = 141. The squares show the distribution of the maxima of individual pulses along the longitude. The designation along the axes is the same as in Fig. 2.} \label{fig:profAB} 
\end{figure}

\begin{figure}
\centering
\includegraphics[angle=0, width=80mm]{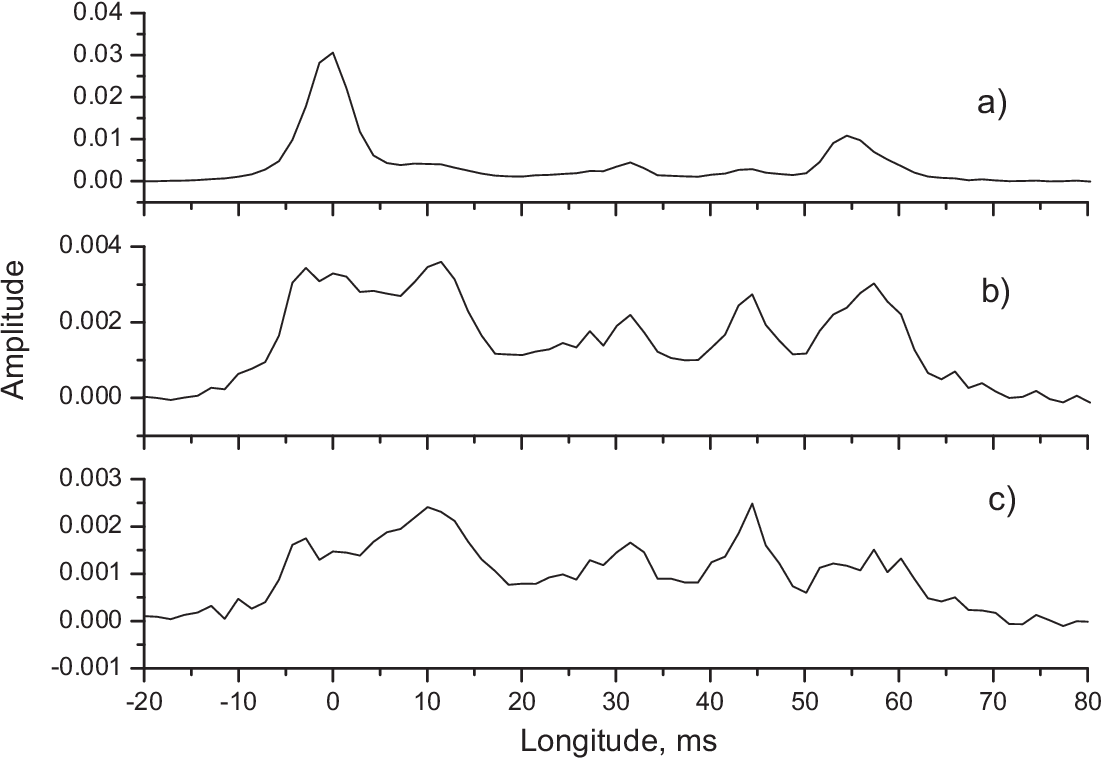}
\caption{Accumulated profiles of pulses in a limited range of amplitudes: a) the average profile for the session on 25.02.24; b) the accumulated profile of 94 pulses with amplitudes ranging from 3$\sigma_N$ to 30$\sigma_N$; c) the accumulated profile of 61 pulses with amplitudes ranging from 3$\sigma_N$ to 15$\sigma_N$. The designation along the axes is the same as in Fig. 2.} \label{fig:profAV} 
\end{figure}
about 0.1 amplitude of the total profile. This is a typical behavior in the QN mode, when about 70\% of the pulses have radiation at all longitudes and their amplitude is significantly less than the amplitudes of strong pulses at extreme longitudes (Fig. 2). In the first and last components, strong pulses are realized, while at other longitudes there is either no radiation at all or very weak. Figure 6 shows, as an example, individual strong pulses over 2 sessions with the QN mode. Their amplitude exceeds the amplitude of the average profile  by 10 times (16.02.24) and by 14 and 16 times for pulses 76 and 35 (25.02.24).
\begin{figure*}[ht]
\includegraphics[width=100mm]{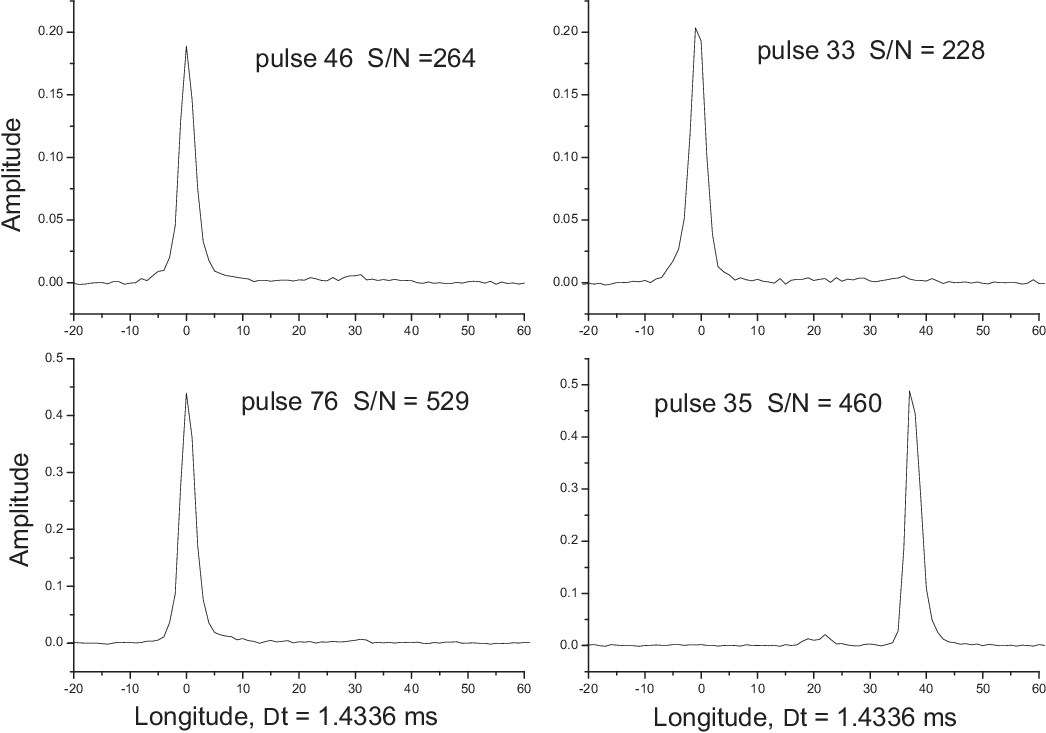}
\caption{At the top – 2 individual pulses for 16.02.24 (QN mode), at the bottom – 2 individual pulses for 25.02.24 (QN mode). The designation along the axes is the same as in Fig. 2.} \label{fig:Single} 
\end{figure*}
 \subsection{Fluctuation spectra}
 Fluctuation spectra provide a good idea of the brightness distribution of individual pulses over longitude and time (see ~\citep{TH1971}).
Figure 7 shows the fluctuation spectra constructed from our observations for three pulsar emission modes with a longitude resolution of 5 ms. The lower figure reflects the session for 25.08.24, corresponding to the quiet normal QN mode. The upper figure refers to the anomalous AB mode (session 16.08.24), and the middle one shows the fluctuation spectrum for the normal enhanced FN radiation mode (session 07/25/23).
 When constructing the fluctuation spectra, 16 longitude intervals with a width of 5 ms each were identified.
 The average profile is shown to the left of each fluctuation spectrum, so that each spectrum can be compared with the corresponding
longitude (point) of the pulse profile. The lower spectrum corresponds to the first point of the average profile for all these figures. The fluctuation spectra were smoothed over three adjacent harmonics to reduce amplitude variations.\\
\begin{figure}[ht]
\includegraphics[angle=0,height=55mm,keepaspectratio]{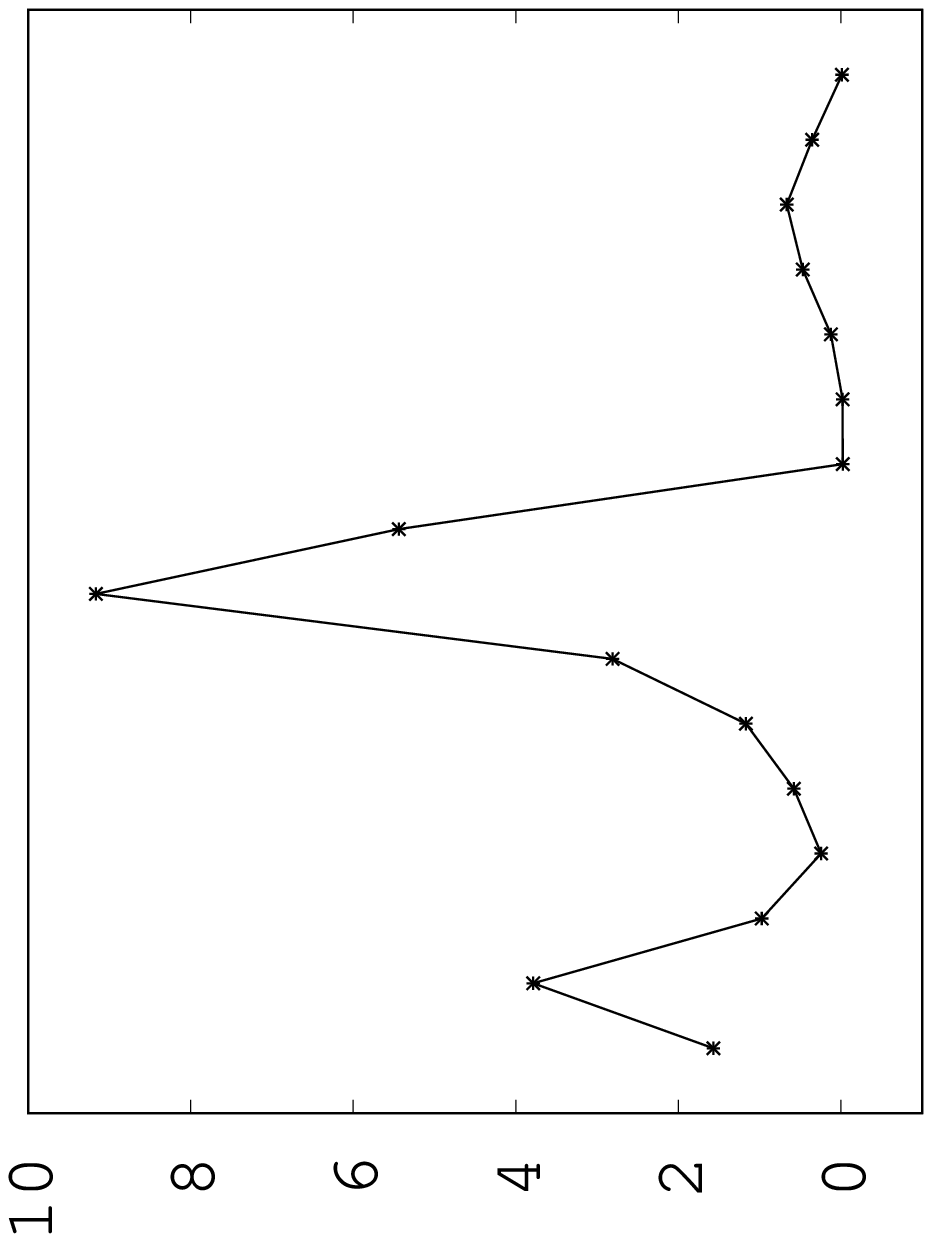}
\includegraphics[angle=0,height=52mm,keepaspectratio]{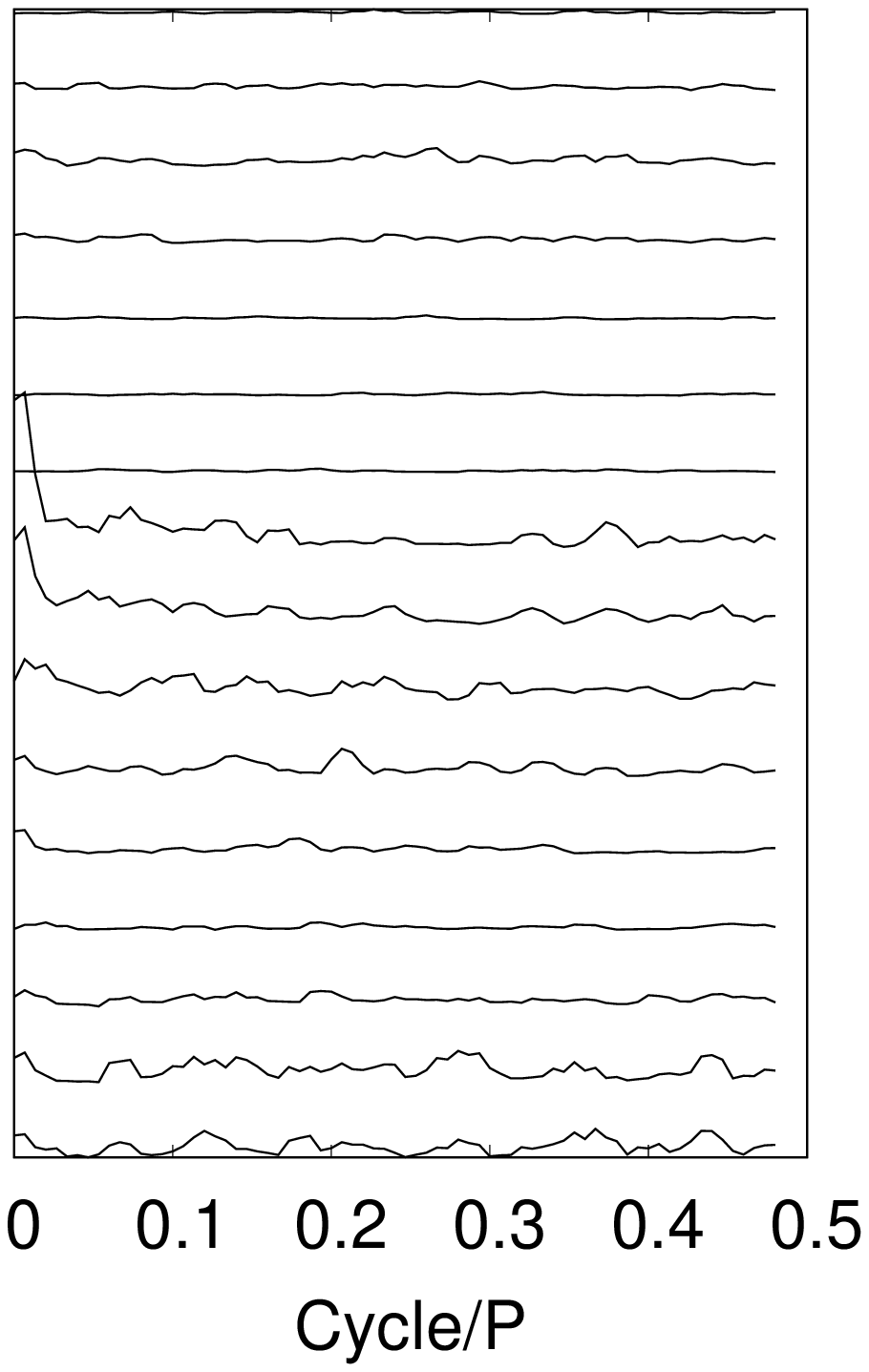}
\includegraphics[angle=0, height=55mm,keepaspectratio]{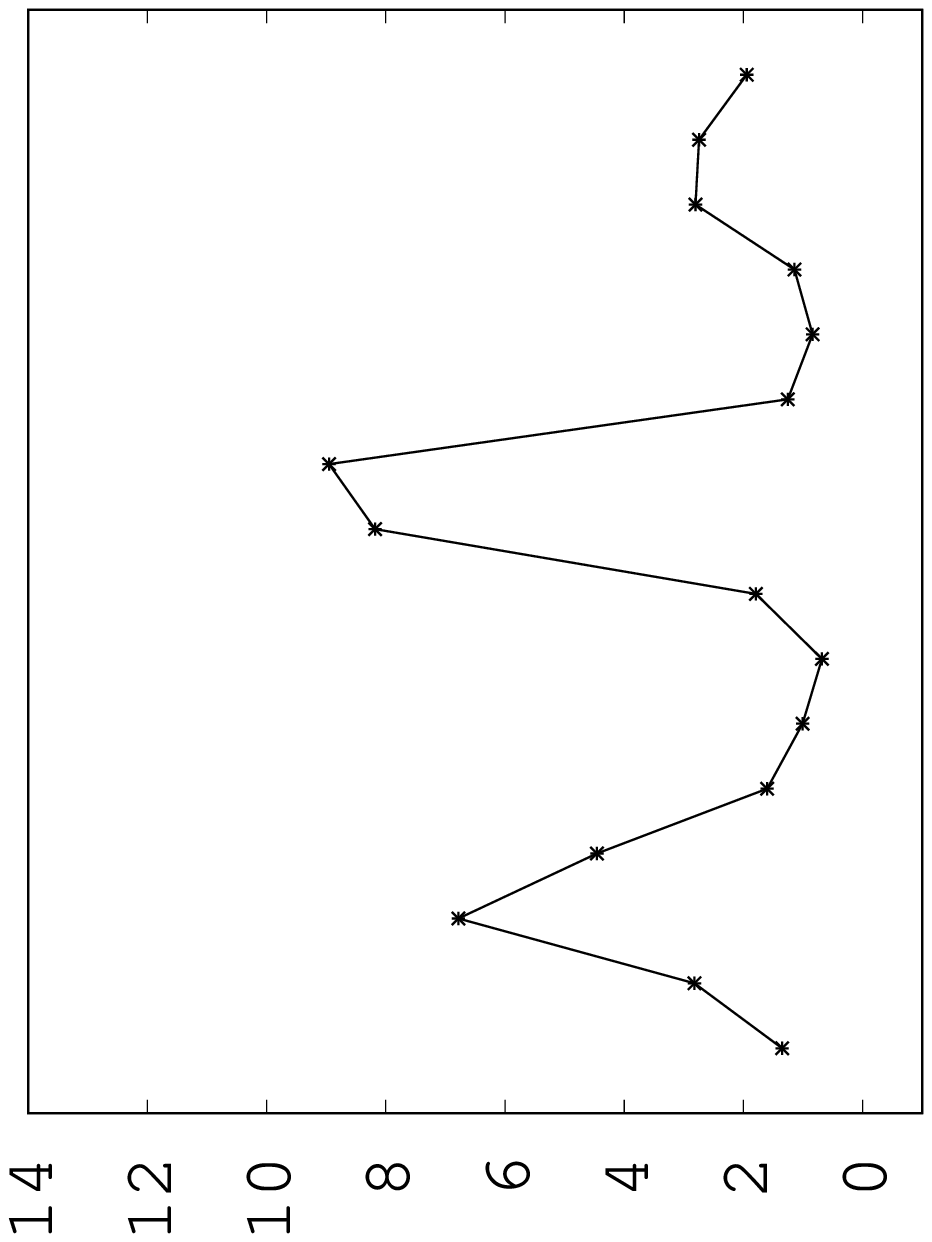}
\includegraphics[angle=0, height=52mm,keepaspectratio]{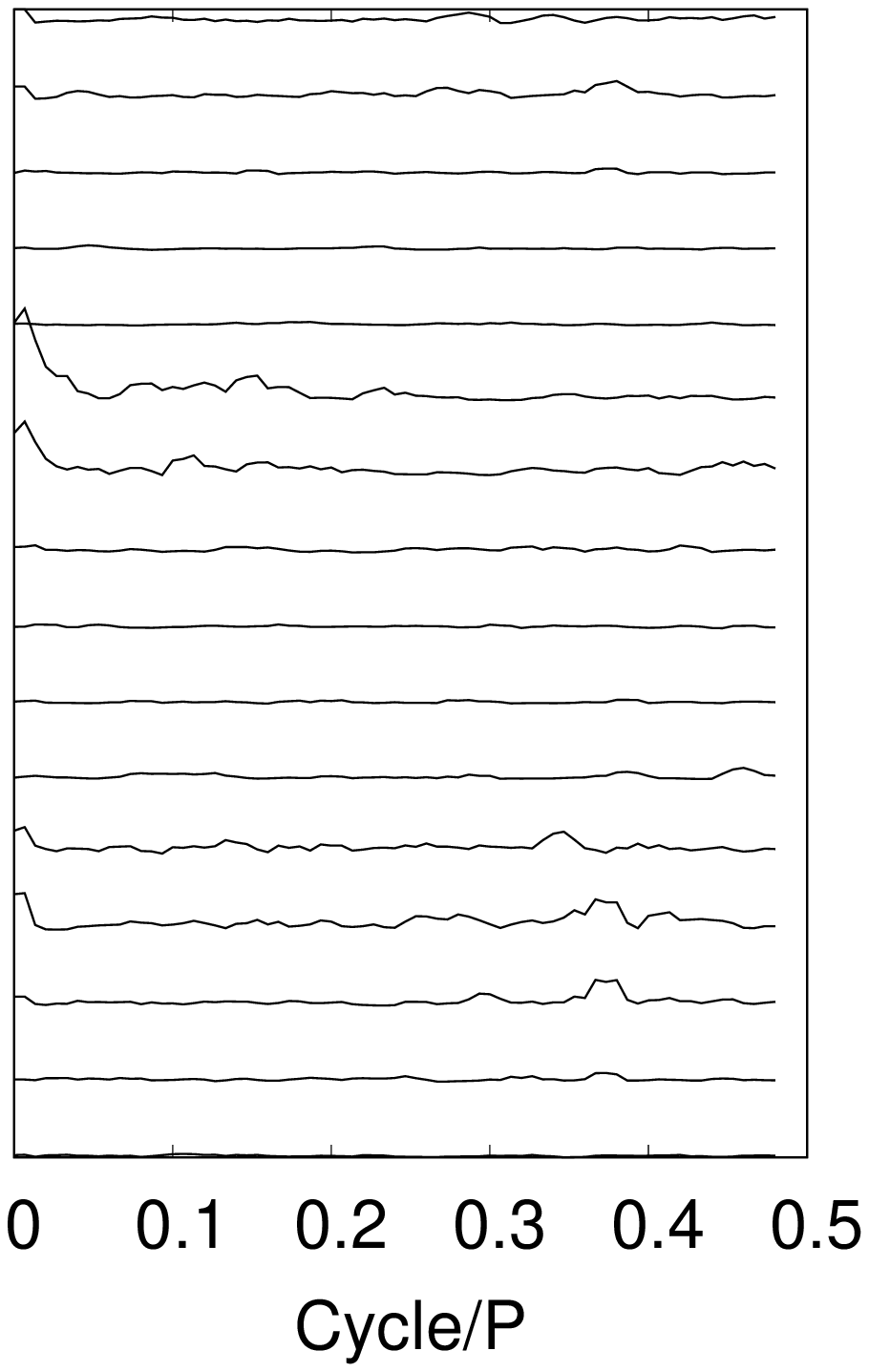}
\includegraphics[angle=0, height=55mm,keepaspectratio]{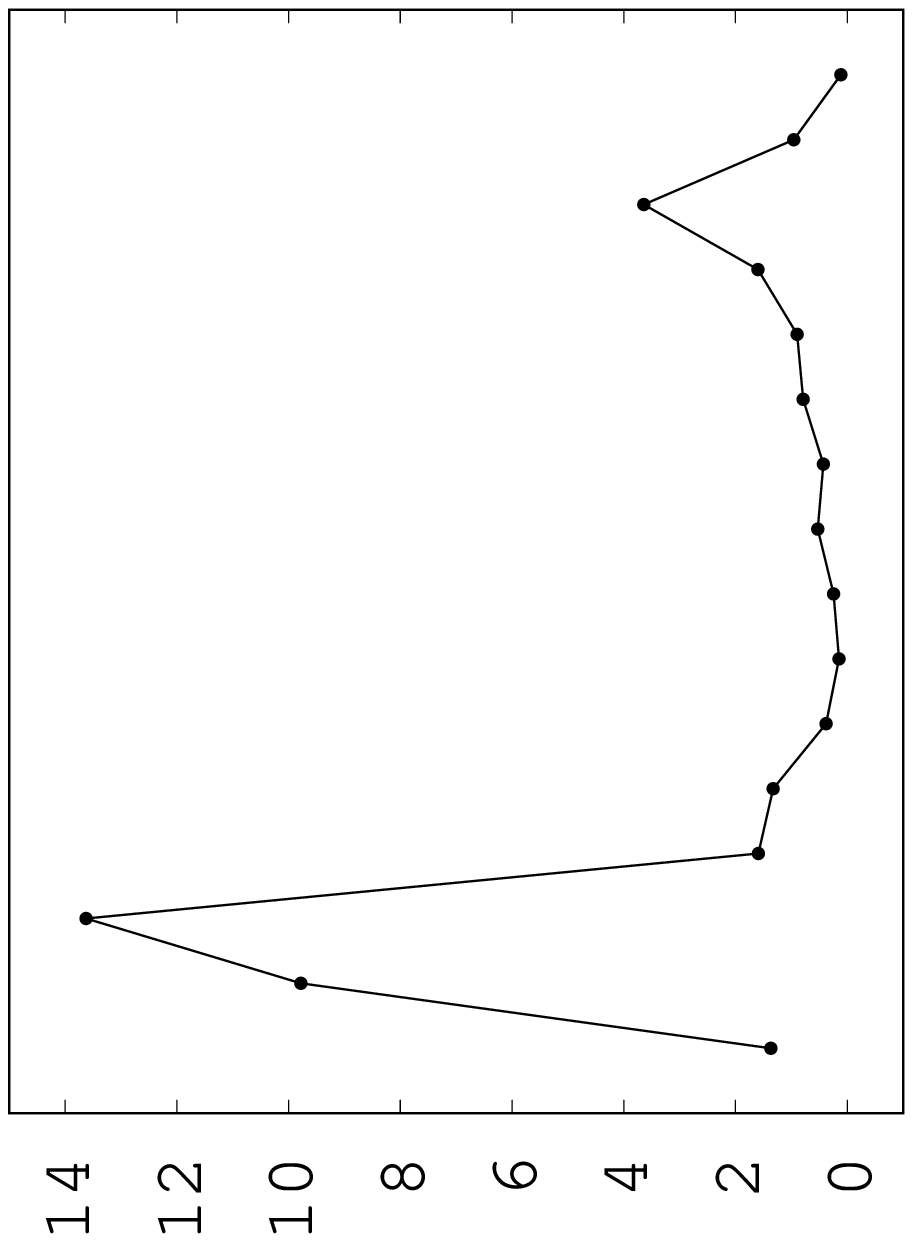}
\includegraphics[angle=0, height=52mm,keepaspectratio]{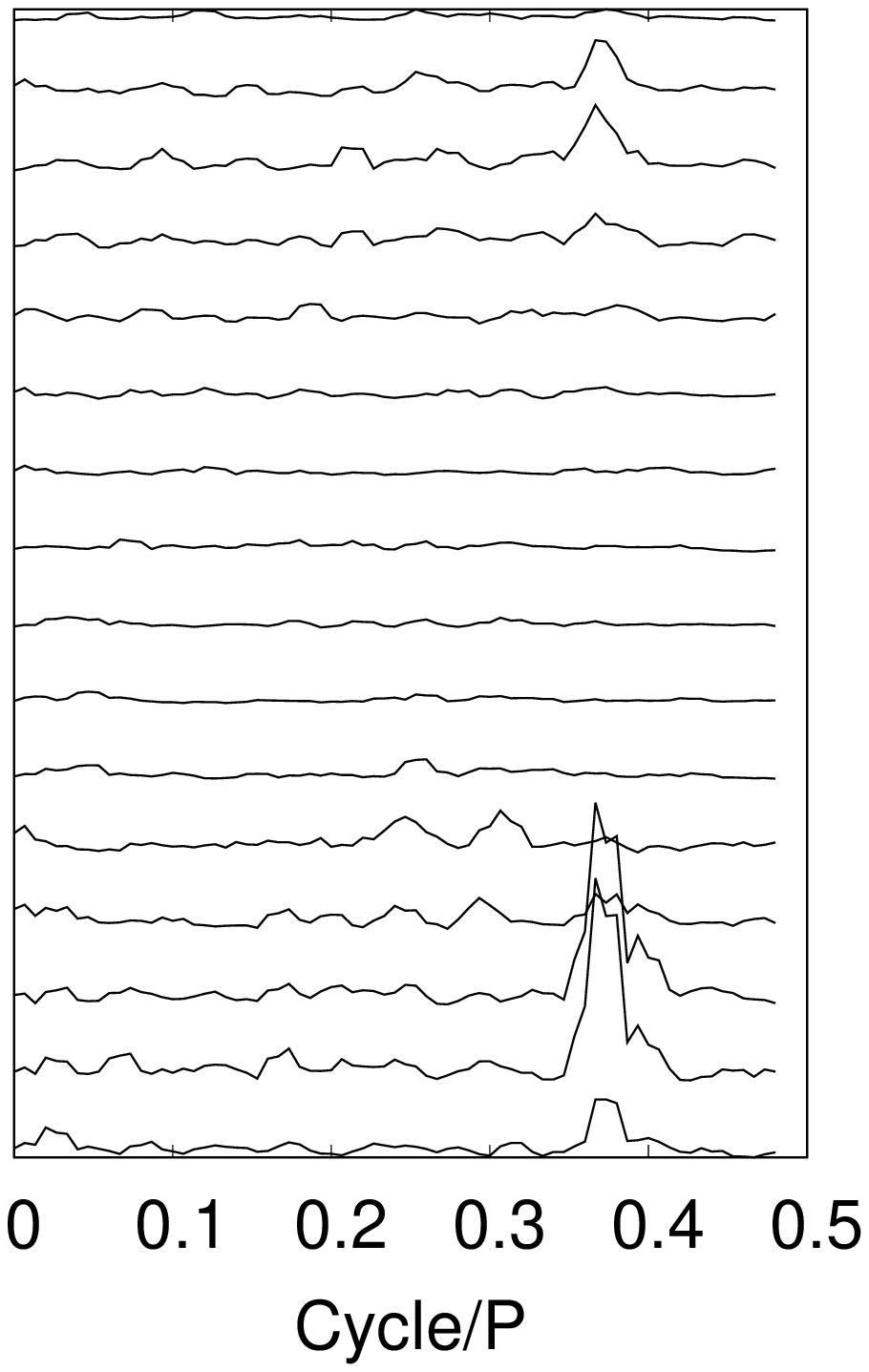}
\caption{Fluctuation spectra for three sessions corresponding to different modes of radio emission. The lower figure represents the session for 25.08.24, corresponding to the quiet normal QN mode. The upper figure refers to the anomalous AB mode (session 16.08.24), and the fluctuation spectrum for the normal enhanced FN radiation mode is shown in the middle (session 25.07.23). The average profile is shown to the left of each fluctuation spectrum. The longitude resolution is 5 ms. The lower spectrum corresponds to the first point of the average profile.}
\label{fig:fluctAB} 
\end{figure}

 A sufficiently compact spectral detail corresponding to the value of
 $P_3{=}2.7P_1$, is clearly visible on the fluctuation spectrum for the QN mode
 at the longitudes of the first and last components of the average profile. 
 This detail can be seen with a greatly reduced amplitude in
the FN radiation mode, and it is completely absent in the anomalous
AB radiation mode. At the central longitudes of the average profile for the FN and AB radiation modes, an increased spectral amplitude is observed at the lowest frequencies, which corresponds to the characteristic time of intensity variations of 10-20
pulsar periods $P_1$; this value can be compared with the period of the carousel received in ~\citep{2014ApJ...792..130M}.

  It is noteworthy that the fluctuation detail corresponding
to the subpulses drift with a period of $P_3{=}2.7P_1$ does not appear in components 2 and 5 (in our classification), i.e. in the components of the inner cone of radiation. This can also be seen in the illustrations provided in publications~\citep{2014ApJ...792..130M},\citep{backer1973}, \citep{1990SvA....34..382P}. A discussion of this feature will be presented in section 4.

  Cross-correlation analysis of amplitude variations at different longitudes has shown that there is no correlation between the 1st and 2nd, 5th and 6th components, i.e. between the inner and outer cones of radiation in the QN and FN modes. There is also no correlation between the extreme and central components for all radiation modes. In some sessions with the QN mode, there is a correlation of amplitude variations between the extreme components with sufficiently large shifts (41 $P_1$ and 29 $P_1$), however, due to the short length of the session, it is impossible to talk about the significance of such correlation.
  
 \subsection{Nullings}
 Along with the manifestation of subpulse drift and a change in the radio emission mode, pulsar B1237+25 has the property of zeroing the pulse intensity for one or more periods - the nulling phenomenon. The work of Rankin~\citep{1986ApJ...301..901R} provides an overview of nulling studies, and the publication by Smith et al.~\citep{2013MNRAS.435.1984S} provides an analysis of nulling parameters for pulsar B1237+25. It was shown that a large proportion of nullings is in the normal quiet mode (QN), and this proportion is 5.2\%, while in the anomalous radiation mode, nullings occur only in 2.7\% of cases.
 
\begin{figure*}
\centering 
\includegraphics[angle=270, width=100mm]{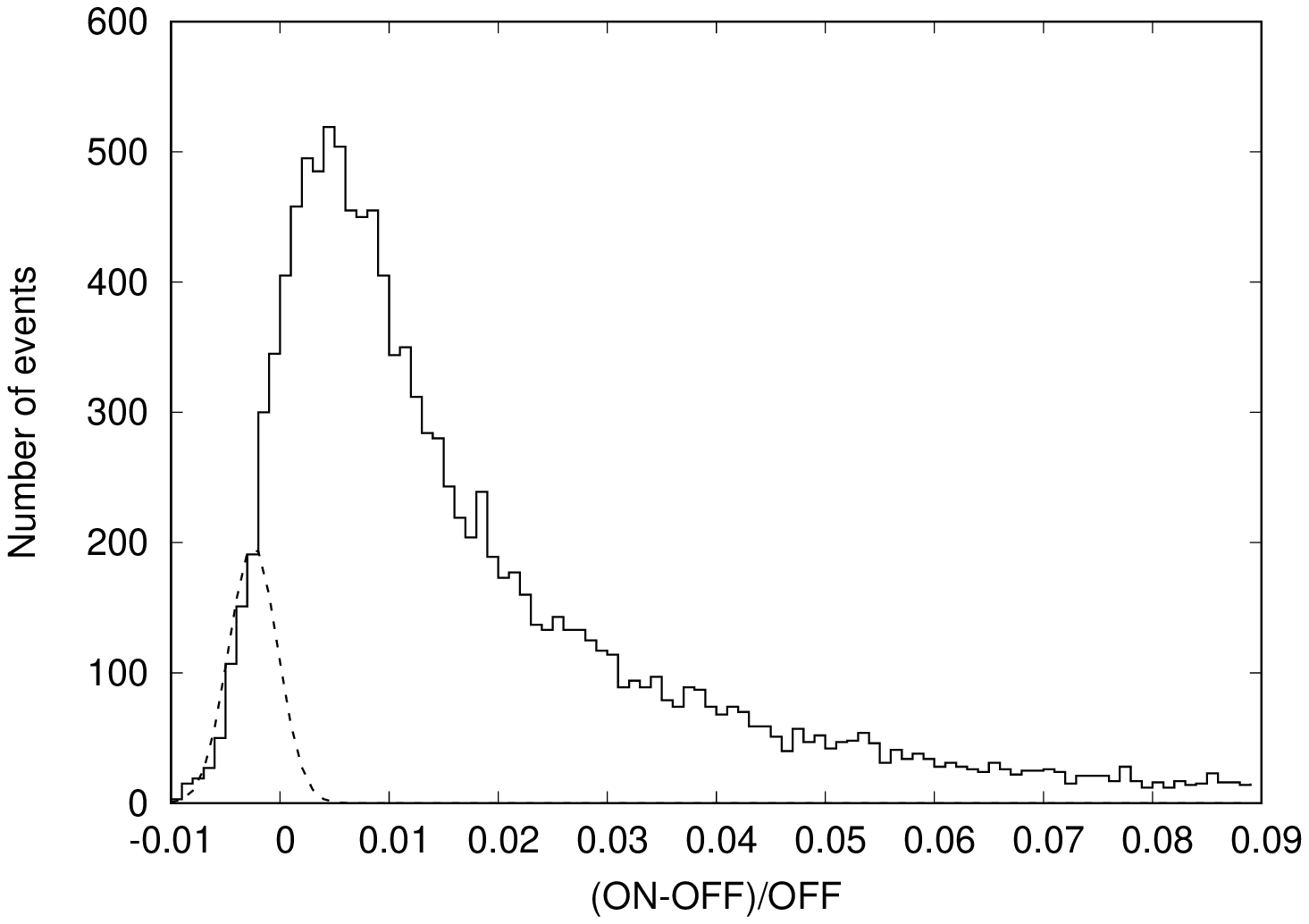}
\caption{Distribution of individual pulses by the magnitude of the relative signal increment in the emission window (ON-OFF)/OFF. The dotted line represents the hypothetical distribution of the excess of weak pulses, which we interpret as nullings.} 
\label{fig:nul} 
\end{figure*}

 We checked the presence of zero pulses
regardless of the radiation mode due to insufficient statistics. Figure ~\ref{fig:nul} shows the distribution of individual pulses according to the magnitude of the relative signal increment in the radio emission window (ON-OFF)/OFF. The dotted line represents a hypothetical distribution of the excess of weak pulses, which we interpret as nullings. The fraction
of such pulses was $8\pm2\%$. This estimate is within the uncertainty limits,
  with the above values, obtained at 327~MHz according to much larger statistics~\citep{2013MNRAS.435.1984S}.

 \subsection{Microstructure of individual pulses}
 When studying the radio emission of pulsars with high temporal resolution, in addition to subpulses, the microstructure of individual pulses with scales from several microseconds to milliseconds is also revealed. Some review of microstructure studies is presented in a recent 
 Popov's publications \citep{popov2024}. This paper describes the technique for
 obtaining microstructure parameters by analyzing autocorrelation functions (ACF). In our observations, the radiation field was recorded
 (voltage) in the frequency band 2.5~MHz, followed by compensation for the effect of radio wave dispersion in the interstellar plasma (see section ~\ref{observe}).
 This provided a time resolution of 0.4~microseconds. The scattering of radio waves by inhomogeneities of the interstellar plasma leads to an uncompensated blurring of the pulses by the amount of the characteristic scattering time $\tau_{sc}$.
 One of the goals of our observations was to determine scattering parameters by measuring the characteristic decorrelation band of spectral distortions,
 caused by interstellar scattering. $\Delta\nu$ varied irregularly in the range of 10-100~kHz, which corresponds to the scattering time 
 2 - 20~microseconds. A discussion of the results of patrolling scattering parameters will be presented in a separate publication.
   \begin{figure*}
\centering 
\includegraphics[angle=270, width=100mm]{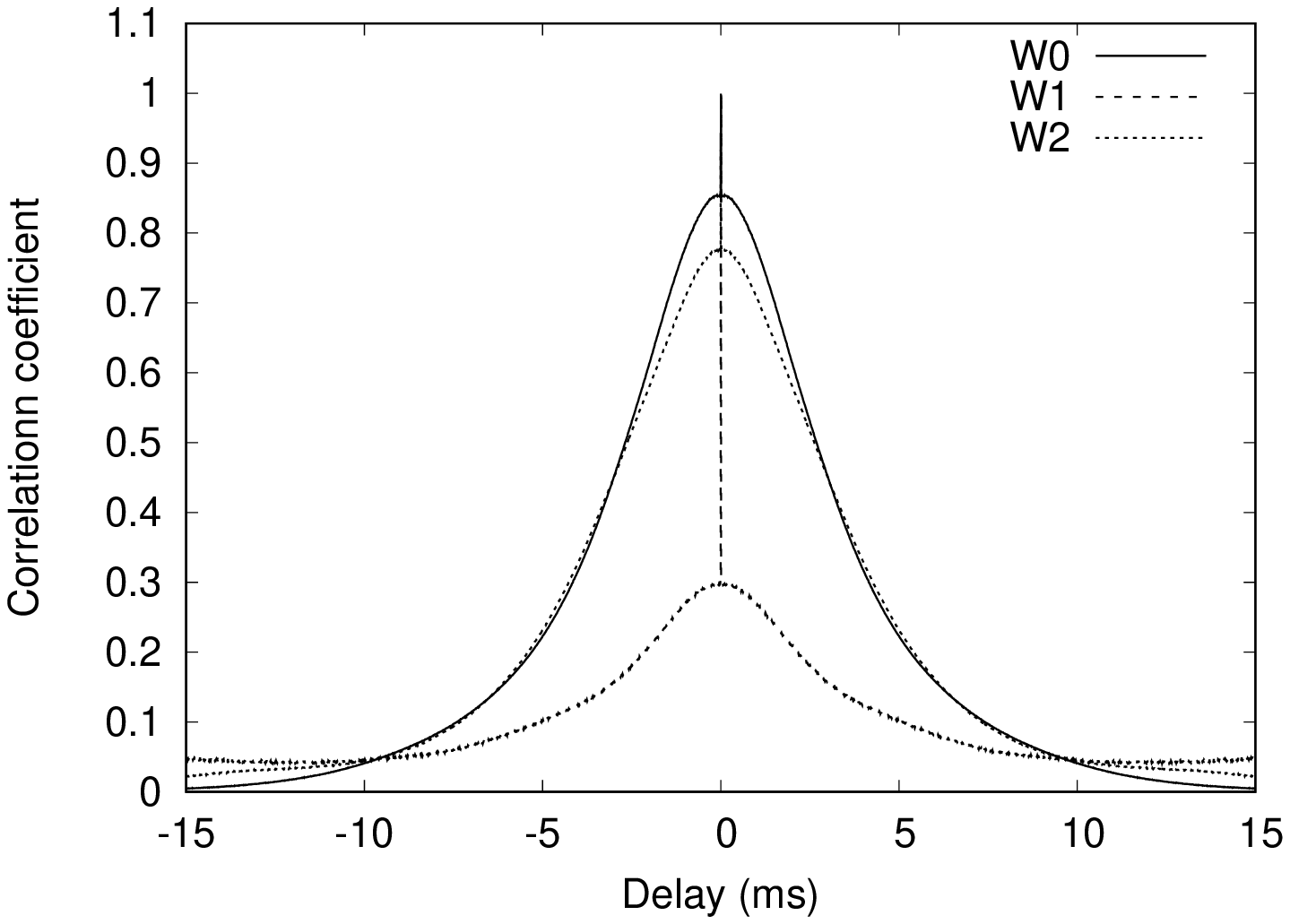}
\includegraphics[angle=270, width=80mm]{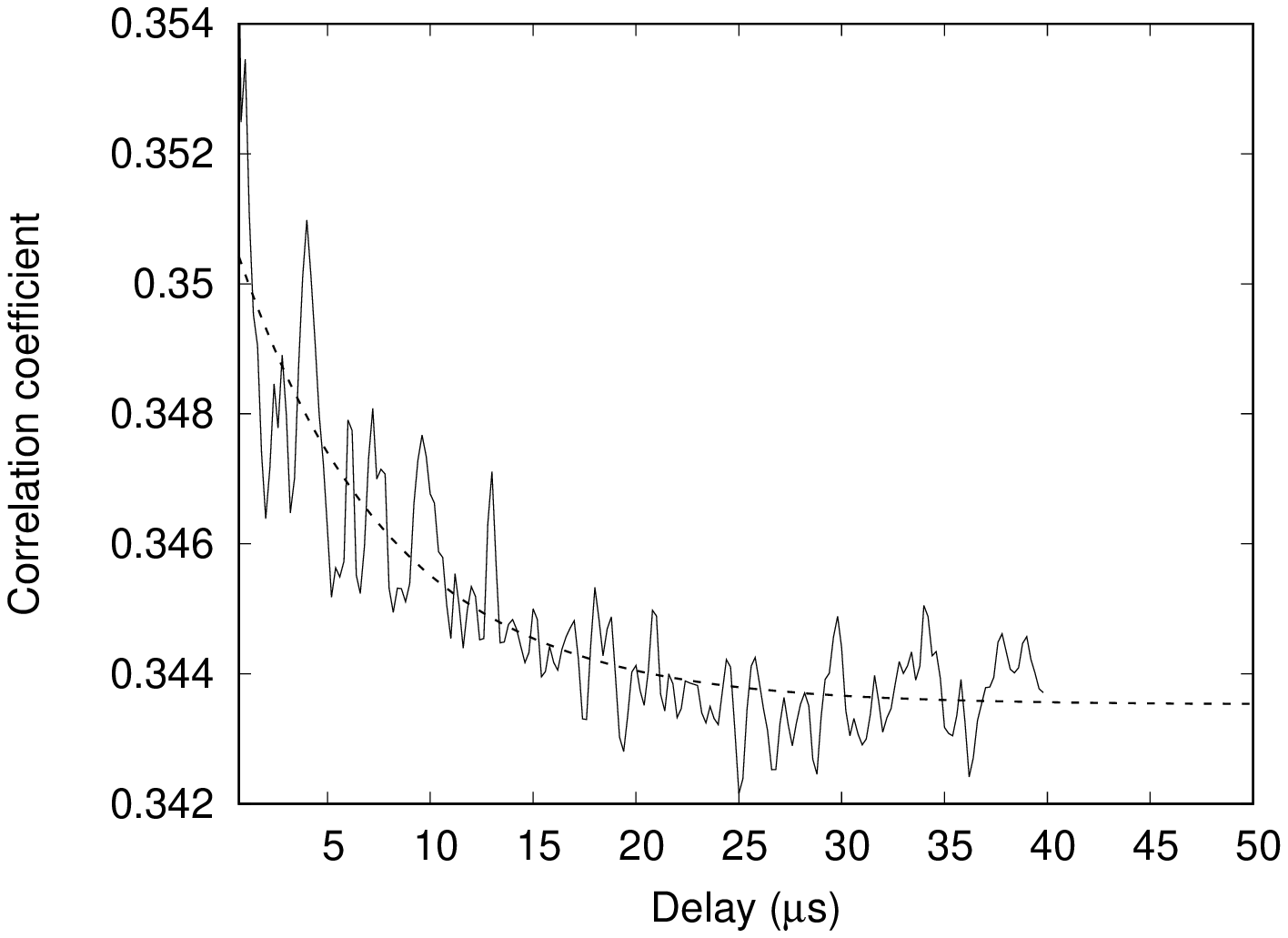}
\includegraphics[angle=270, width=80mm]{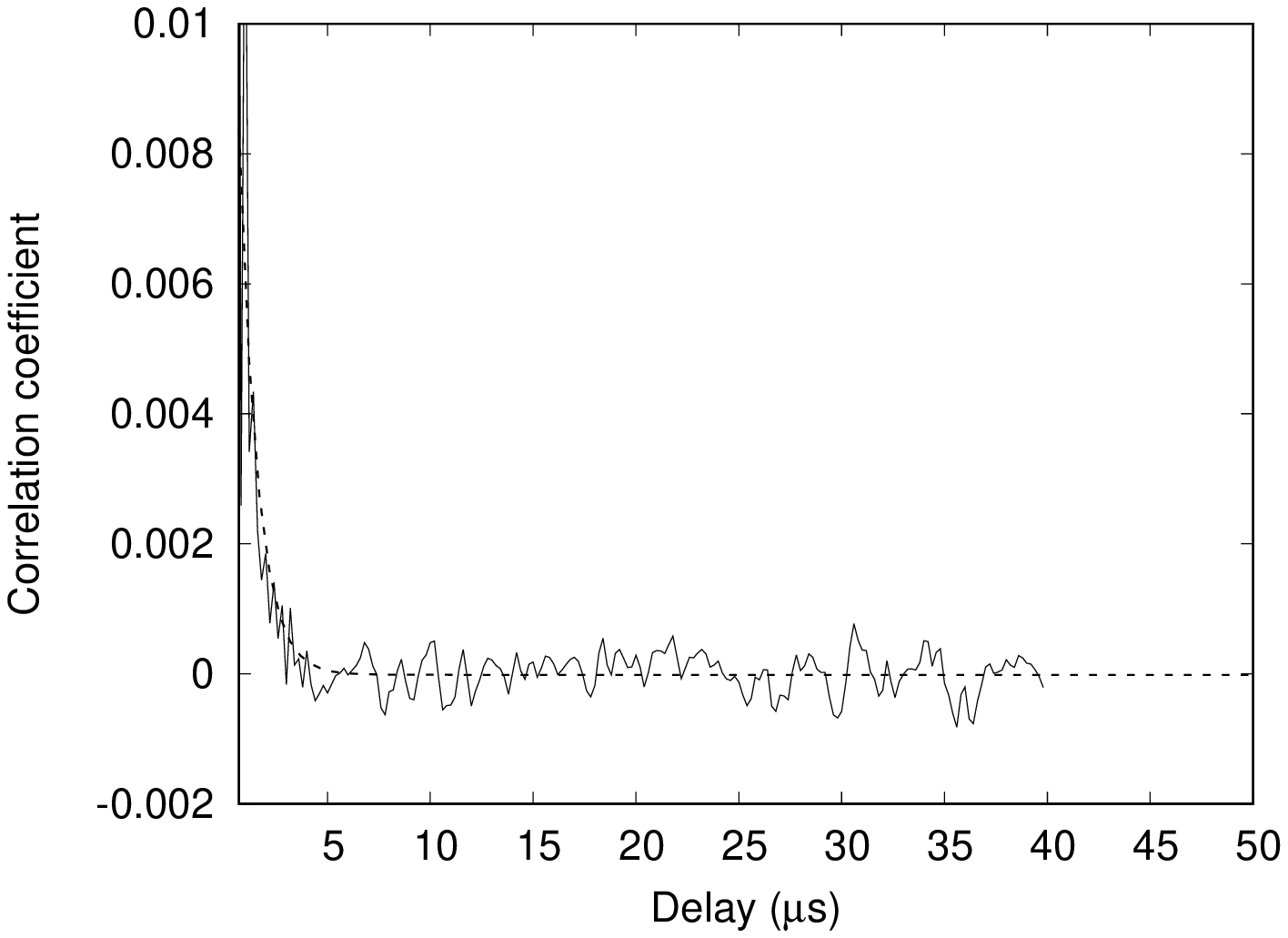}
\caption{Average ACFs obtained by averaging the ACFs
of individual pulses. The average ACFs for three longitude intervals W0, W1, and W2 are shown above. The initial sections of the average ACF
with high time resolution $\delta t{=}0.2\mu s$ are shown below. 
 The left figure represents the average ACF on the pulse (ON), and the right one corresponds to the recording noise areas (OFF). See the explanations in the text of the article.} 
\label{fig:mic} 
\end{figure*}
 Figure ~\ref{fig:mic} shows the average ACF obtained by averaging the ACF
of individual pulses. The upper figure shows the ACFs for three longitude intervals W0, W1 and W2; the interval
W0 (-10~ms - 20~ms) corresponds to components 1 and 2 (the front of the cone), the interval W1 (20-40~ms)
 corresponds to the central components 3 and 4, and the interval W2 (40-70~ms) corresponds to components 5 and 6 (tail of the cone). These ACFs were calculated for a signal averaged over 50 points, i.e. with a time resolution of 10 microseconds. There are no signs of microstructure in the average ACF at all longitudes. Half-widths of
 the sub-pulses were $3.03\pm0.05$~ms, $2.70\pm0.06$~ms, and $3.34\pm0.05$~ms for the longitude intervals W0, W1, and W2, respectively.

The lower figures show the average ACF for selected strong pulses
 regardless of their longitude; recordings with the maximum temporal resolution of $\delta t{=}0.2 \mu s$ were used to construct these ACFs. 
 Pulses whose amplitude exceed the $15\sigma$ level were selected.  There were 220 such pulses in total; the sample was taken from 16 sessions in which
 a high level of pulsar radiation was observed. For each selected pulse, the ACF was calculated for the ON recording section in the range of $\pm1$~ms (10,000 samples) from the burst maximum. For the same pulse, the control section of the OFF noise recording was selected. The figure~\ref{fig:mic} at the bottom
left shows the initial section of the ACF in the pulsar radiation window ON,
and on the right shows the initial section of the ACF on noise OFF. The initial part of the ACF at zero shift, where the values are one, and with neighboring shifts equal to $0.2\mu s$ and $0.4\mu s$ are not shown in the figures to maintain the desired scale. The lower right figure shows the receiver's transfer function, which reflects the shape of the band.

  The initial section of the average ACF for the signal (left figure) is radically different from the receiver transfer function. There is a sequence of bursts of decreasing amplitudes with
a characteristic time of $\leq 1$~microseconds. The dotted line corresponds to
 an exponential function with a time constant of $7.5\pm 0.5$~microseconds, which was obtained by data approximation. The dotted line in the right figure corresponds to an exponent with a time constant of $0.9\pm 0.2$~microseconds.
 The value of 7.5~microseconds is in good agreement with the expected scattering time. The resulting scattering time corresponds to the decorrelation scale of 21.2 kHz.  Thus, we have discovered a microstructure with a characteristic time
of $\leq 1$~microseconds, which is manifested in the characteristic width of bursts on the ACF.
  \begin{figure*}[ht]
\begin{center} 
\includegraphics[angle=270, width=80mm]{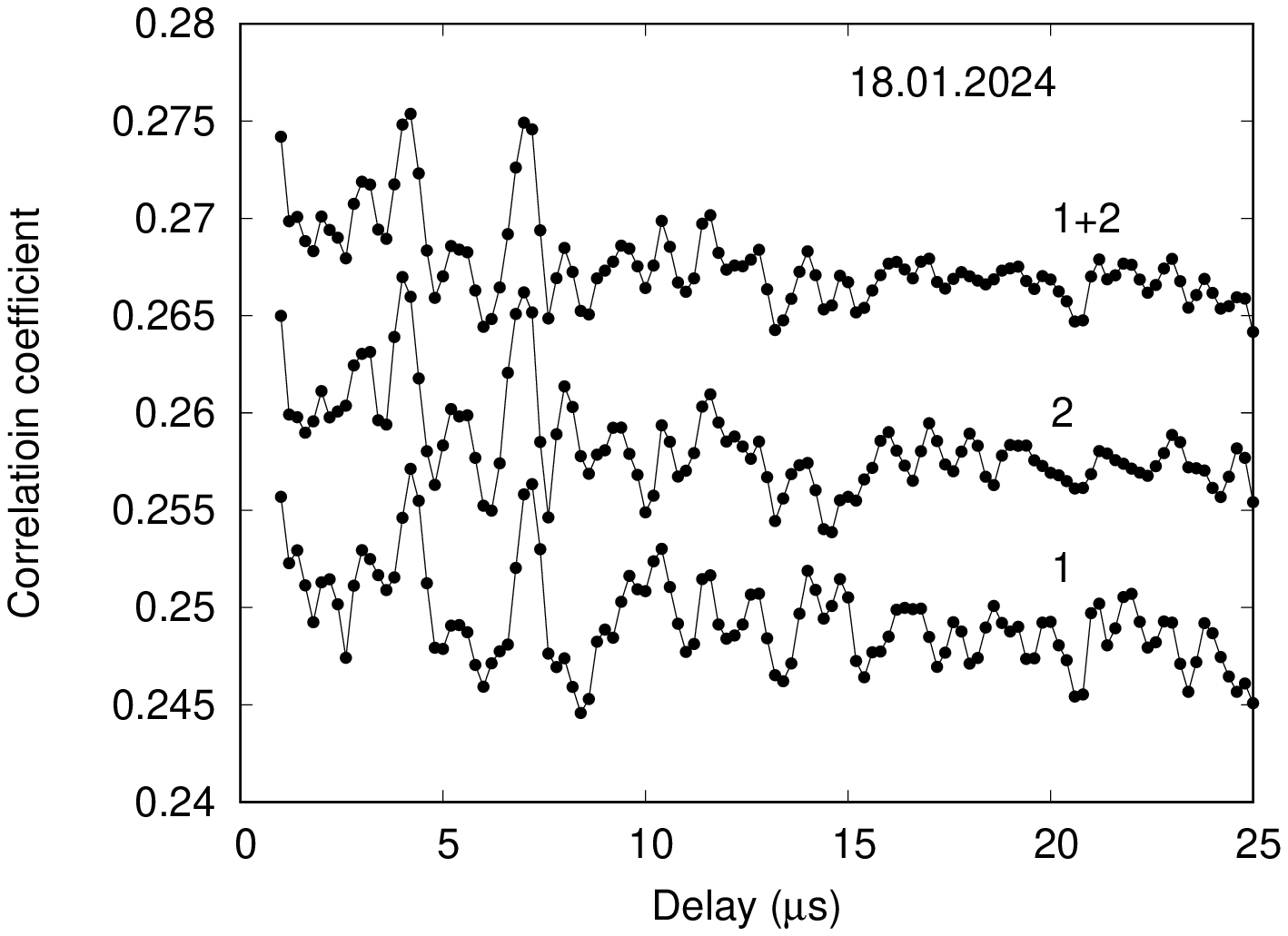}
\includegraphics[angle=270, width=80mm]{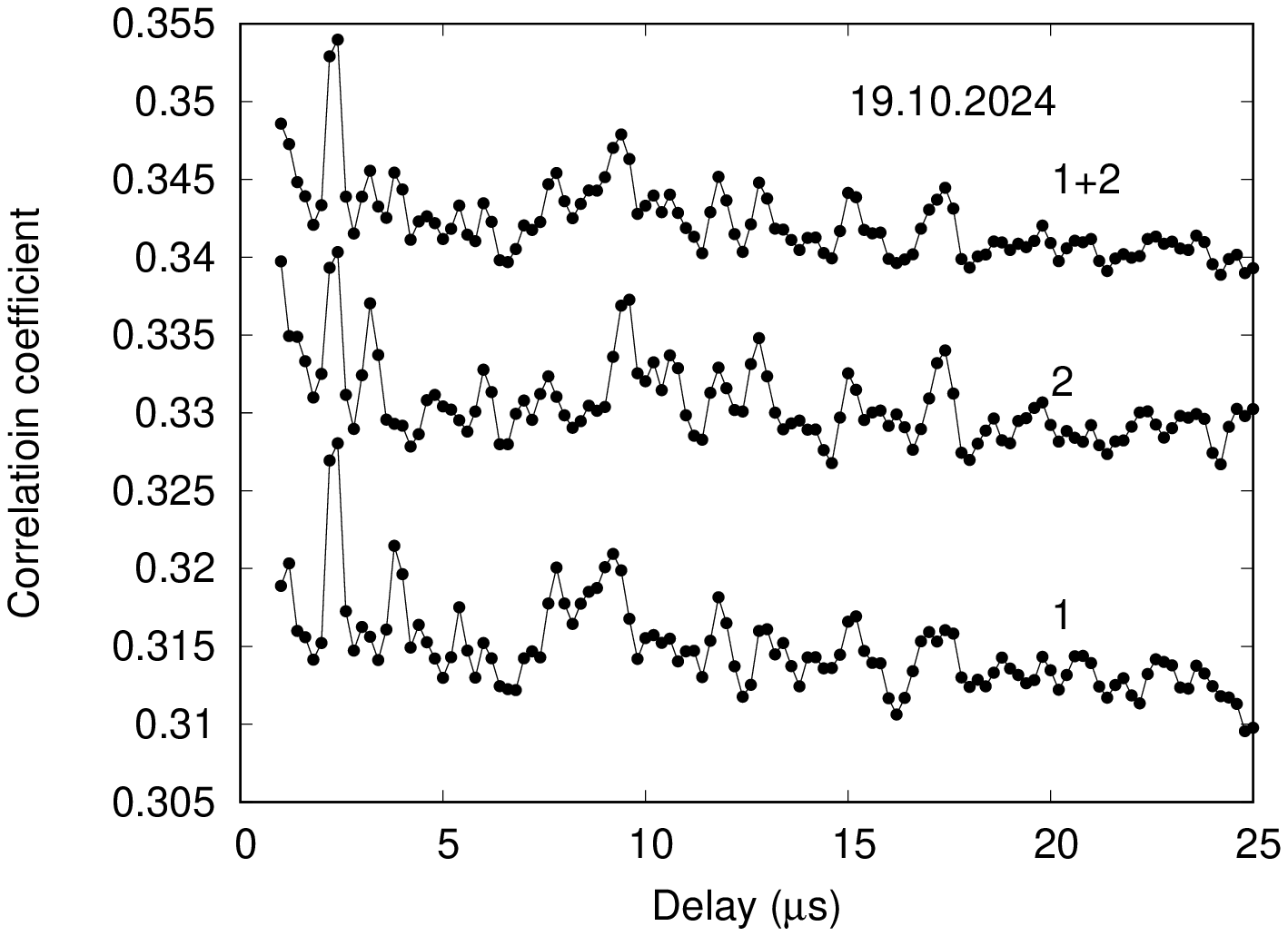}
\includegraphics[angle=270, width=80mm]{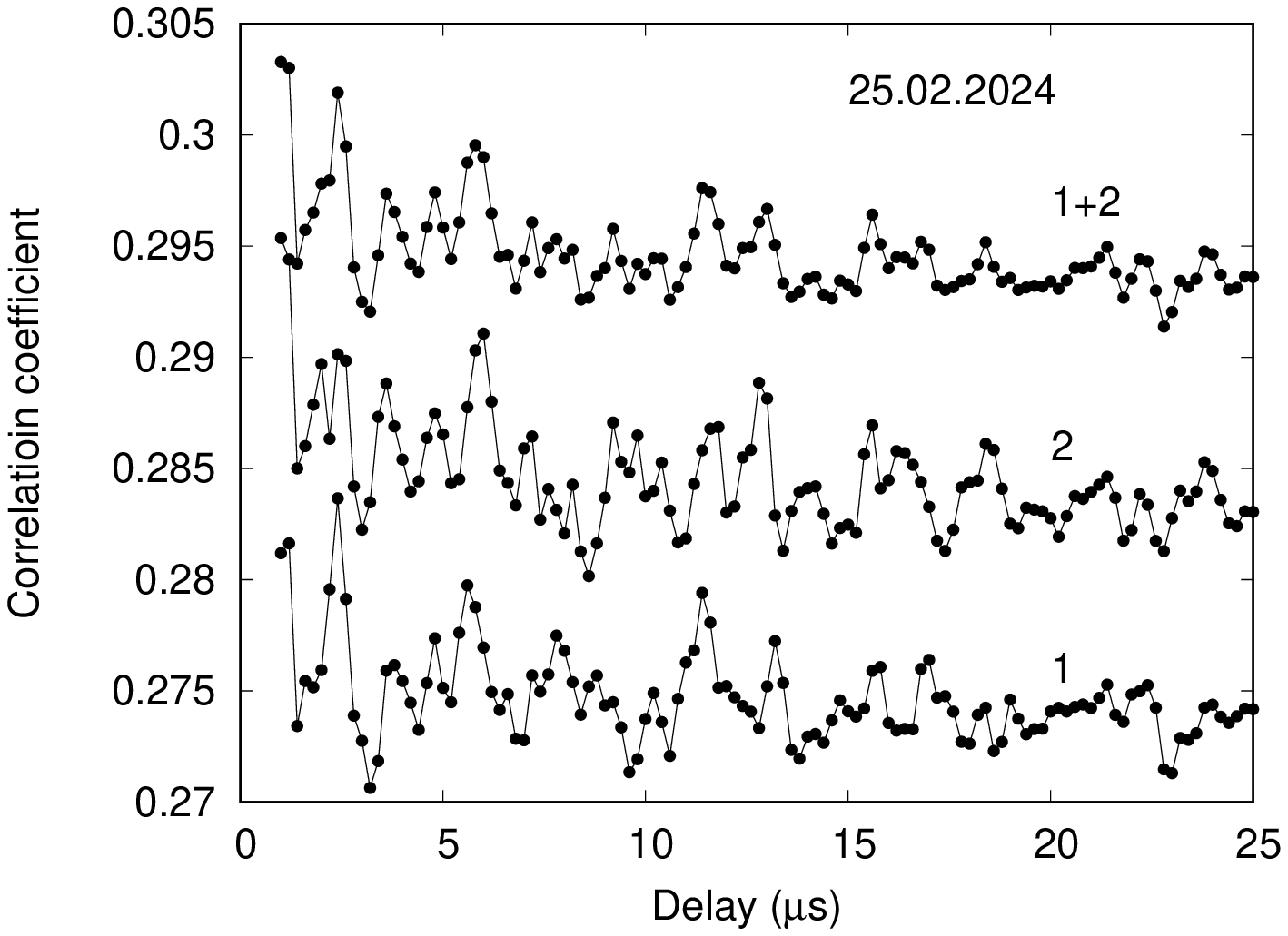}
\includegraphics[angle=270, width=80mm]{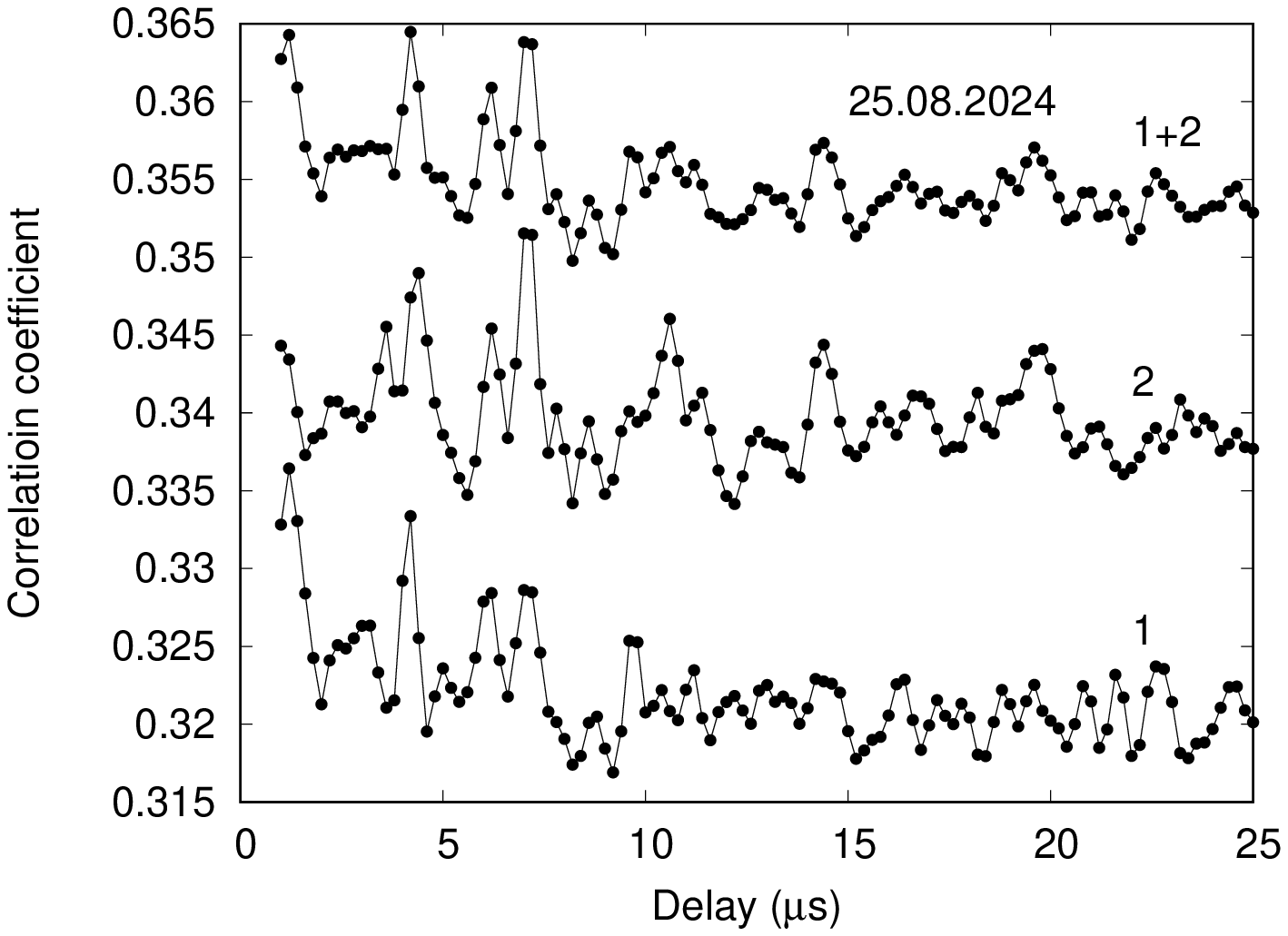}
\caption{The initial ACF sections, averaged for selected observation dates, with each session divided into two parts, indicated in the figures by numbers 1 and 2; the upper graph shows ACF for the entire session (1+2). Each drawing also shows the date of the observation session.} 
\label{fig:scat} 
\end{center}
\end{figure*} 

To confirm the presence of micropulses manifested in the characteristic structure
of the initial ACF section, average ACFs were constructed for two independent samples of strong pulses in separate observation sessions. The results are shown in figure~\ref{fig:scat}. The ACFs of such independent samples are indicated in each figure
by the numbers 1 and 2, and the upper graph corresponds to the total ACF for the entire observation session (1+2). About 20 strong pulses were selected in each session, respectively.
 The fraction contained about a dozen pulses. It can be seen that the structure of the initial section of the average ACF for independent pulse samples agrees well, but differs for different sessions.
 This confirms the assumption that this structure reflects
the geometry of the delay paths of micropulses as a result of the scattering process. 
 The characteristic true duration of these hypothetical micropulses should be less than 0.5 microseconds, which is manifested in the small width of the modulation bursts at the initial ACF section.

 \section{Discussion\label{disc}}
The purpose of our research is to obtain from observations characteristics that are useful for theoretical interpretation.
Pulsar B1237+25 provides a unique opportunity to study
the properties of radio emission throughout the polar cap, as the line of sight passes very close to the magnetic pole; the deviation is $0.25^{\circ}$, or $2.5\%$
of the angular radius of the polar cap~\citep{2013MNRAS.435.1984S}. The average pulsar profile has been analyzed and discussed in many publications
(see, for example, \citep{2013MNRAS.435.1984S}). Usually the average profile is divided into 5 components, which are associated with two radiation cones: the outer and inner, and
the central (core) component.

    \begin{figure*}[ht]
\begin{center} 
\includegraphics[angle=270,  width=80mm]{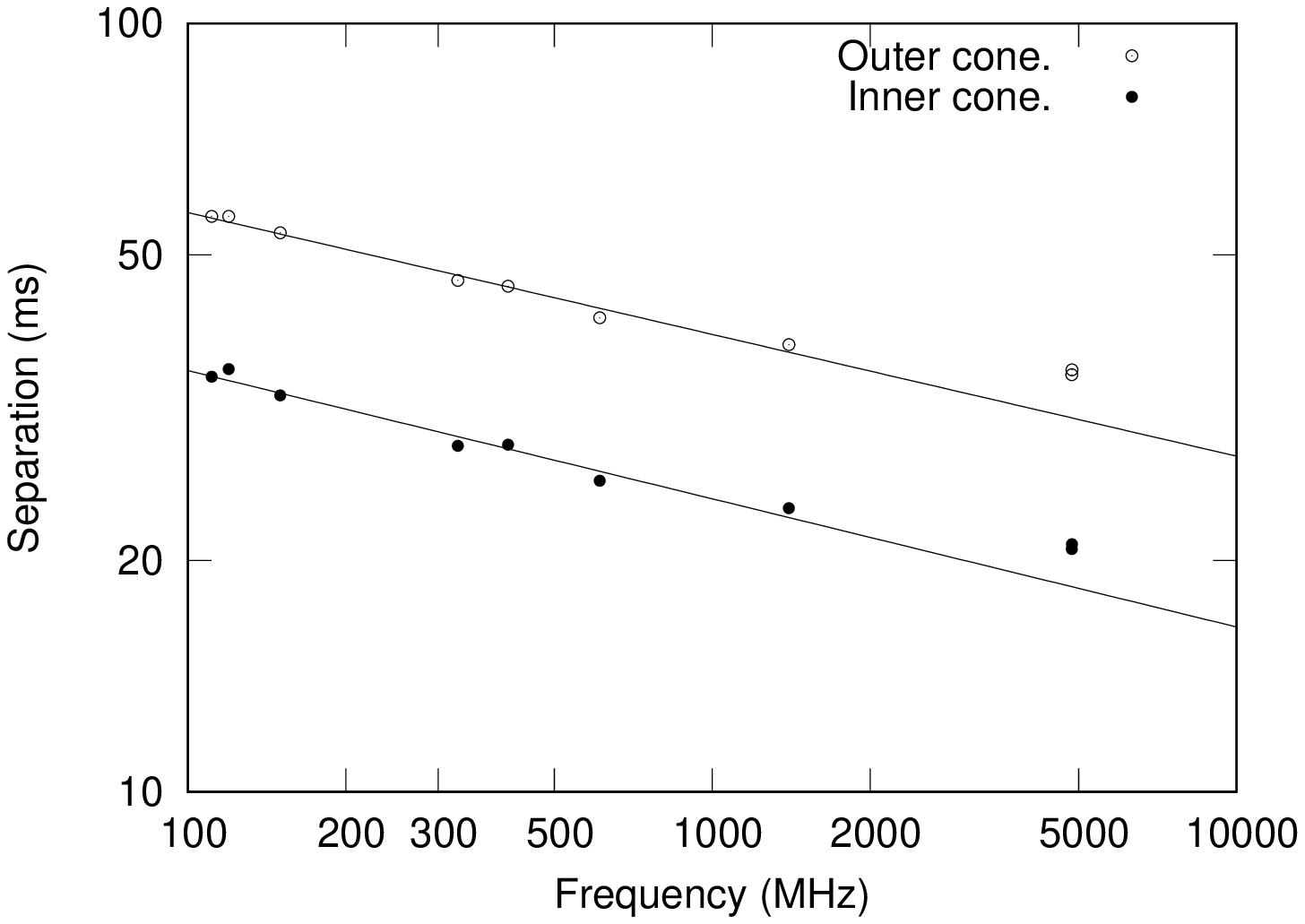}
\includegraphics[angle=270,  width=80mm]{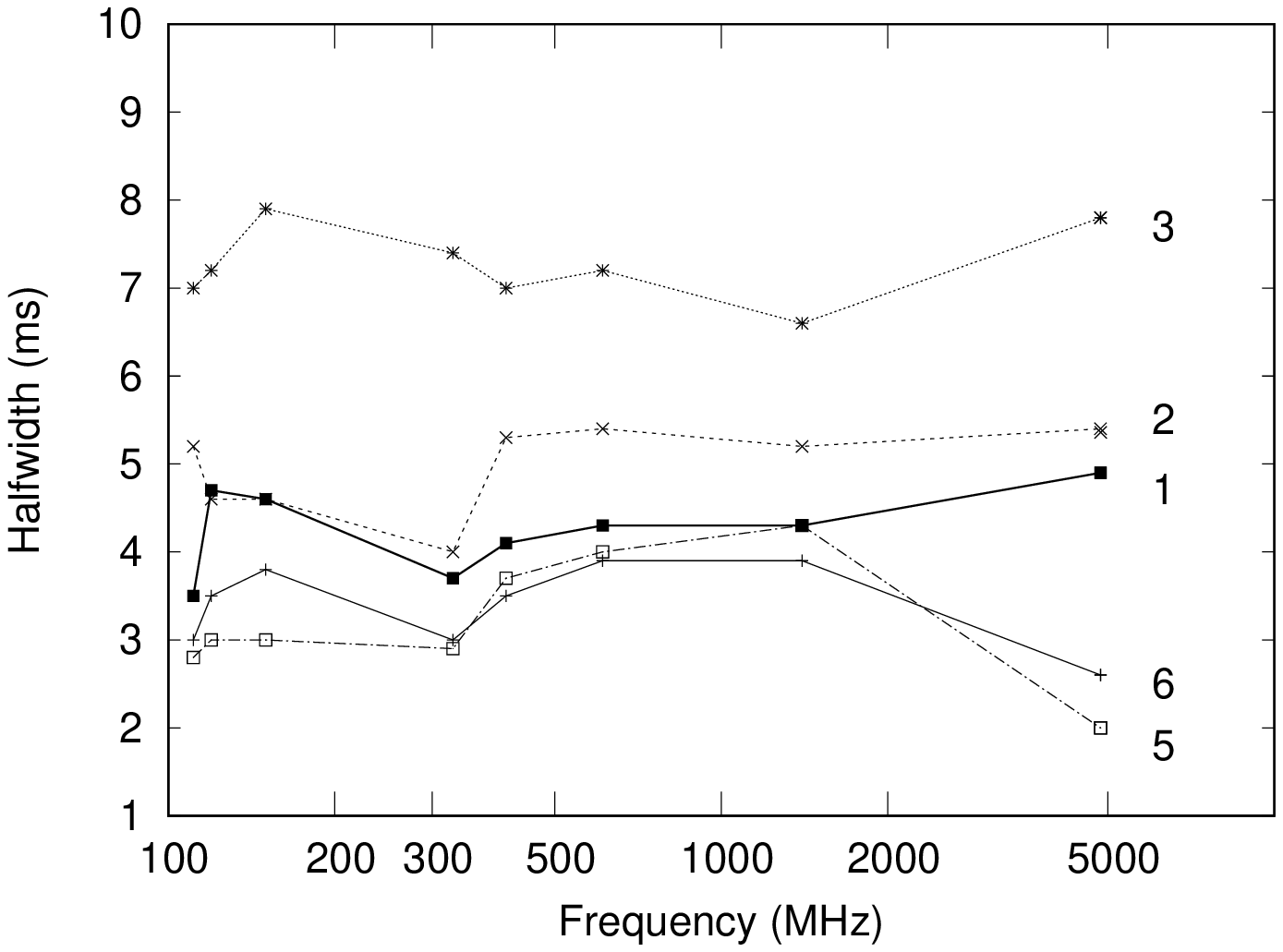}
\caption{Figure 11. Frequency dependence of the distance between the components
belonging to the outer and inner cones of radiation (on the left) and the half-width of the components on the frequency (on the right). The left figure is made on a logarithmic scale, and the right one is on a semi-logarithmic scale; the numbers of the curves in this figure indicate  the component number.} 
\label{fig:separation} 
\end{center}
\end{figure*}

We found at 111~MHz that the central component consists of two components:
a broad 7.0~ms and a compact 1.6~ms; in different modes, these component half-widths are different (see table~\ref{tab:width}).
  The figure~\ref{fig:separation} shows the frequency dependence of
  the distances between the components of the outer and inner cone of radiation. 
  In this case, the characteristics of the average profile combining radiation in all modes were used. Data on other frequencies are taken from the following publications: 139.3~MHz and 150.8~MHz~\citep{2016A&A...591A.134B},
  327~MHz~\citep{2019MNRAS.489.1543O}, 408~MHz and 610~MHz ~\citep{1998MNRAS.301..235G}, 1400 
  ~MHz~\citep{2018MNRAS.474.4629J,1999ApJS..121..171W}, 4850~MHz
  ~\citep{1997A&AS..126..121V}. The straight lines in the figure show the approximation of the measured values by a power-law function
  $\Delta\Phi(\nu){=}C\nu^{\alpha}$.  
  Measurements at a frequency of 4850 ~ MHz were not used in the approximation, since it is known that in the studied dependence for many pulsars, a break is observed at frequencies above one gigahertz. The values of the exponent  $\alpha_1{=}-0.158\pm0.005$ and $\alpha_2{=}-0.167\pm0.008$ for the outer and inner cones, respectively, were obtained. Within the limits of uncertainty, these values are the same.
  
  Theoretical approaches to explaining the observed diversity of average pulsar profiles have been presented in recent years by
  Beskin et al.~\citep{andr2010,besk2012MNRAS,hak2014ARep, hak2017MN,besk2023MN}. They considered both the processes of secondary plasma formation and the effects of radio emission propagation in the pulsar's magnetosphere. Satisfactory explanations for complex pulse profiles, including polarization properties, have been demonstrated.
 
  Back in 1988
  Beskin, Gurevich, and Istomin \citep{BGI1988} published basic approaches to the theory of pulsar radio emission. In particular, they showed that in the relativistic
inhomogeneous plasma leaving the pulsar's magnetosphere along
open magnetic field lines, two modes of radio emission are generated: the ordinary mode (O) and the extraordinary mode (X). The extraordinary mode propagates in a straight line, that is, tangentially to the magnetic field lines
in the radio emission generation zone, while the ordinary mode propagates along the magnetic lines to a certain output level.
  Thus, the ordinary radiation mode forms the external components of the average profile. 
  (the cone), and the extraordinary one is the central components (the core).  Beskin, Gurevich and Istomin \citep{BGI1988} obtained
  theoretical estimates of the $\alpha$ index for the frequency variation of the angular dimensions of cones corresponding to the O and X modes. According to these estimates, $\alpha{=}-0.14$ and $\alpha{=}-0.29$ for the inner and outer radius of the cone.
    Thus, an increase in the distance between the components of the average profile was predicted, as well as an expansion of the profile components themselves with a decrease in the frequency of radio emission. 

    Our measurements of $\alpha$  gave the same values
    for components of the outer and inner cones: 
    $\alpha_1\approx\alpha_2{=}-0.16$, 
and this value is in satisfactory agreement with the predictions for the behavior of O-mode radiation. Therefore, we believe that the inner and outer cones of the average profile of the pulsar  
B1237+25 components (1, 2, 5, and 6) represent a single external cone of O-mode radiation, which is separated from the magnetic field line at the same height, determined by the boundary of the open field lines
corresponding to the full width of the average profile.
In our view, components 3 and 4 are the central (core) components of X-mode radiation.
The size of this central core is determined by the width of component 3 and is about
7~ms at all frequencies, as shown in the right panel of the figure~\ref{fig:separation}.

In the fundamental work of Beskin, Gurevich and Istomin~\citep{BGI1988} the parameter Q is introduced, which characterizes the pulsar's "efficiency": namely, the level of electric potential above the polar cap. This parameter is defined by the following expression $Q=2P_1^{11/10}\dot P_1^{-2/5}$, where $P_1$ is the pulsar period in seconds and $\dot P_1$ is the derivative of the period in units of $10^{-15}$~s/s. For pulsar B1237+25, Q=2.6. It is claimed that for pulsars with Q>1, the instability of the process of secondary plasma generation over the polar cap can be predicted. Such pulsars are subject to nulling phenomena, subpulse drift, and radiation mode changes. This is exactly the behavior observed in our pulsar. The subpulse drift, however, is observed in the normal QN radiation mode only in the extreme components 1 and 6. The discharges feeding the secondary plasma in these components cannot freely propagate to the longitudes of the inner cone due to competition with the activity zones at these longitudes. Native activity zones are present at these longitudes in smaller numbers and, apparently, cannot form a structure for a stable carousel. The entire system of active zones is unstable, and the normal QN radiation mode is interrupted by nullings and transitions into the abnormal AB mode. The abnomalous mode reflects another stable configuration of activity zones, about which we only know that: 1) the structure of the activity zones at the edge of the outer cone is destroyed (the carousel stops working); 2) the distance between the inner and outer cones is almost doubled from 8.3 ~ ms to 16.3 ~ ms (see Table 1), and the distance between the inner cone and the core is reduced and filled with radiation. Apparently, at these longitudes, the discharges propagate from the point of origin to the magnetic pole. A significant amplification of the compact component 4 is understandable, since, in our opinion, this component integrates radiation from the entire ring of the nearest activity zones.

We will follow the reasoning presented in Popov's work~\citep{popov2024}.
This interpretation uses the traditional 
  hollow cone model (Radhakrishnan and Cook \citep{Rad1969}). According to the total angular width of the outer cone and the central cone, it is possible to estimate, respectively, the height above the polar cap of separation from the direction of the magnetic field lines of radiation of the ordinary mode (O) and the height of radiation generation of the extraordinary mode (X).
 The equation for the last closed field line in polar coordinates
looks like this:
  \begin{equation}     \label{eq:dipole}
  r(\phi)=R_Lsin^2\phi,
\end{equation} 
where $R_L$ is the radius of the light cylinder ($R_L{=}P_1c/2\pi$), and the angle $\phi$ is calculated
from the direction of the magnetic axis. The full width for the O and X components can be obtained from the parameters of the average profile shown in Table~1. For this purpose,
the half-width of the extreme components (1, 6 and 3, 4) will be taken at the 1/10 level, and the tabular values should be multiplied by the coefficient $k=\sqrt{(ln10/ln2)}=1.82$.  These values should be added to the angular distance between components 1, 6 and 3, 4, respectively; in this case, the desired angle $\psi$ will be equal to half the full opening of the hollow cone. The obtained angle values correspond to the direction of the tangents to the magnetic field lines in the radiation generation region, and the angle $\phi$ for substitution into formula 1 is one and a half times less \cite{Pop1990}.
It is also necessary to adjust the multiplier $sin^2\zeta$ to account for the angle $\zeta$ between the magnetic axis and the axis of rotation.
Subject to these amendments 
 the final formula for estimating the level of radio emission output in the magnetosphere for the outer cone ($R_{out}$) is:
  \begin{equation}     \label{eq:Rout}
  R_{out}=R_Lsin^2[2/3(\psi/2+1.82w_{1/2})]sin^2\zeta,
\end{equation}
where $R_L$ is the radius of the light cylinder 
($R_L{=}P_1c/2\pi{=}6.6{\times}10^4$~km), $\psi$ is the angular
distance between the maxima of the extreme components of the average profile (1.6), and $w_{1/2}$ is the average half-width of these components; for the central edge component (3) only the half-width of the component remains in the formula.
 The following values of the height above the polar cap were obtained for the X-mode radiation generation zone and for the O-mode  heights
of 80~km and 370~km, respectively; these values are $0.1\%$ and $0.8\%$ of the radius of the light cylinder for pulsar B1237+25 at a frequency of 111~MHz.

We explain the splitting of the radiation cone into two components by the competition
of the vacuum gap discharge regions in the polar cap. A similar model was considered by Gil and Sandick~\citep{2000ApJ...541..351G}. As mentioned earlier, Beskin~\citep{1982SvA....26..443B} showed that electrical discharges
generating secondary plasma
shield each other and cannot be located closer than a certain distance $r_p$ from each other. This distance is approximately equal to the height
of the vacuum gap $h_p$. The most productive discharges occur at the border
cones of open magnetic field lines, where the radius of curvature of these lines $\rho$ is the smallest and, accordingly, the shortest free path during the development of an avalanche ~\citep{1982SvA....26..443B}.
   \begin{figure*}[ht]
\begin{center} 
\includegraphics[angle=270, width=120mm]{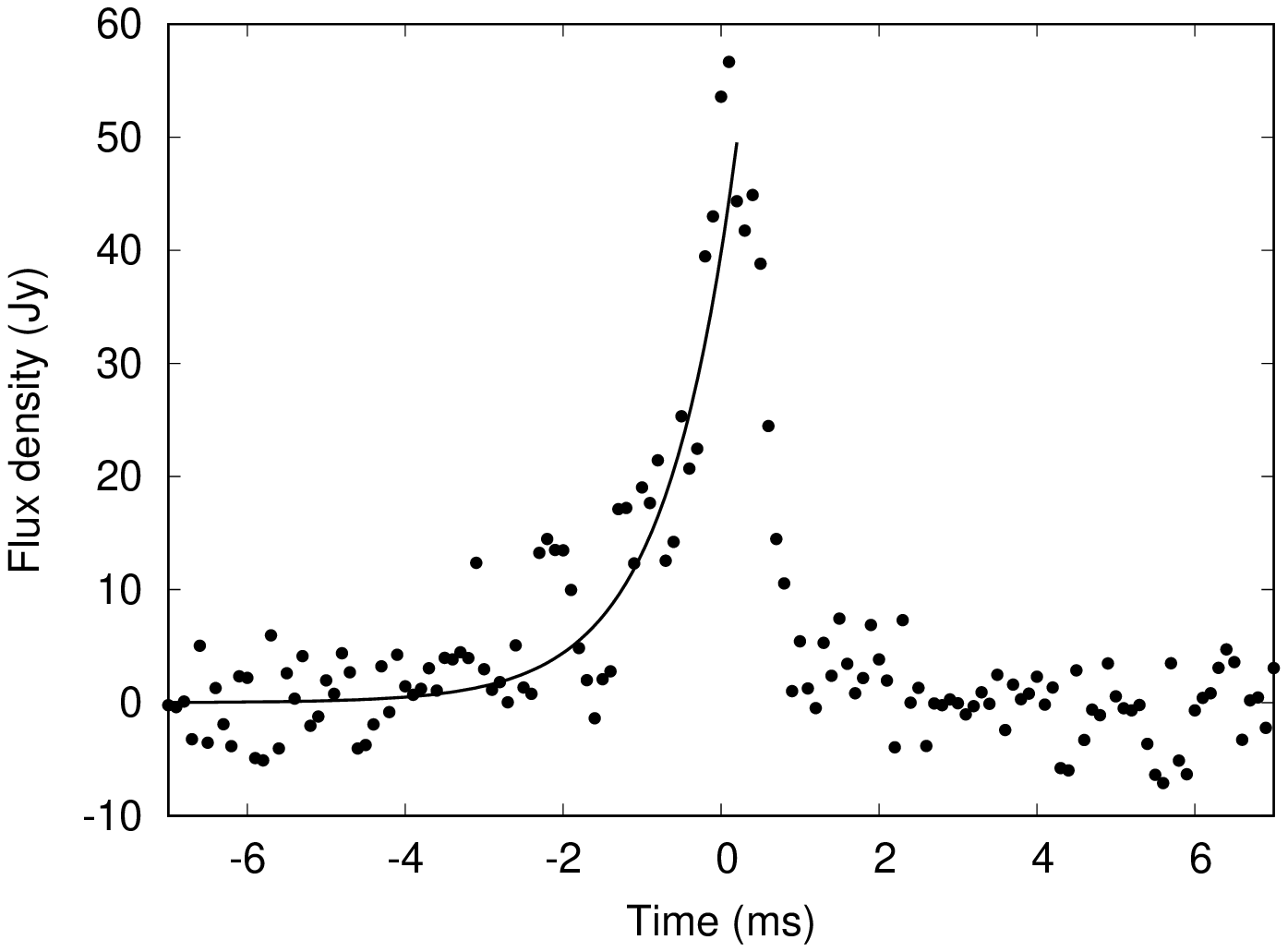}
\caption{An individual pulse belonging to the first component of the average profile. This example illustrates the avalanche-like process of secondary plasma formation. The solid line corresponds to an exponent with a time constant of 1 ms.} 
\label{fig:separate} 
\end{center}
\end{figure*}
 Maan et al.~\citep{2014ApJ...792..130M} attempted  to construct a scheme for the distribution of activity spots in the outer cone of pulsar B1237+25. According to their estimates, a set of 18 equidistant activity zones is formed at the boundary of the hollow cone.
 The activity zones (spots) rotate near the magnetic pole in crossed ($\textbf{E}\times\textbf{B}$) electric and magnetic fields, forming a so-called carousel. Maan et al.~\citep{2014ApJ...792..130M} consider that
the period of circulation of the carousel is $28.4P_1$.
A thread of discharge spark that originated at some distance from the magnetic pole
in the polar cap, will rapidly shift to the magnetic pole along the magnetic meridian at a velocity
of approximately 1~km/s~\citep{1982SvA....26..443B}.
The evolution of a spark discharge that is accompanied by the avalanche-like birth of secondary plasma
is a complex unsteady dynamic process~\citep{besk2023MN}. 

We will try to identify some characteristic parameters of this process
that are necessary for interpreting the results of our research.
The characteristic discharge development time is $\tau_0{=}\Lambda h_p/c \approx 1$~microseconds, where $\Lambda$ is a multiplicity factor of approximately 20~\citep{1982SvA....26..443B}. The excitation zone (spot) has
certain dimensions in the direction opposite to the magnetic meridian.
One of the mechanisms of discharge spark spreading, is called the "photon burst" effect, was proposed by Cheng and Ruderman~\citep{1977ApJ...214..598C}. It consists of a very energetic secondary electron (positron) hitting the surface of the pulsar, and as a result, a secondary high-energy ($E {>}$ 1~MeV) gamma quantum is emitted. This radiation is non-directional, and such a photon burst can effectively expand the spark in almost all
directions within the range of the free path
of the gamma quantum before the next generation of an electron-positron pair.
Thus, the fully developed region of secondary plasma birth 
also has a certain width in the direction perpendicular to the magnetic meridians.

For our pulsar, the line of sight passes very close to the magnetic pole
($\beta{=}-0.3^{\circ}$). Therefore, it is possible to analyze the discharge development along the magnetic meridian. It is immediately clear
that the discharge that begins at the boundary of the polar cap never develops to the magnetic pole (see Figure~\ref{fig:Single}).
This also follows from the characteristic width of the subpulses,
which is about 3~ms (see section ~3).
Such attenuation of the secondary plasma formation process may be due to saturation of the avalanche-like plasma generation process or
competition with other areas of activity in the polar cap (Beskin shielding effect~\citep{1982SvA....26..443B}). Figure ~\ref{fig:separate} shows an individual pulse belonging to the first component of the average profile. It shows exponential growth and a sharp break with a front duration of less than one millisecond. 
Such a sharp drop in intensity corresponds to the angular magnitude of the pulsar rotation of less than $\omega{=}4.5\times 10^{-3}$~rad. This allows us to obtain a restriction on the $\gamma$ factor of the relativistic secondary plasma responsible for radio emission: 
$\gamma{>}1/(\omega sin\zeta){=} 260$~\cite{Bartel2002}; here $\zeta{=}57^{\circ}.6$
is the angle between the magnetic axis and the rotation axis of the pulsar~\cite{2013MNRAS.435.1984S}.

In Subsection 3.5, evidence of the presence of a microstructure with a characteristic scale $\tau_\mu\le0.5$~microseconds was presented. For this value $\tau_\mu$, the height of the vacuum gap $h_p\le750$ is ~cm. 
A microstructure with such a short time scale has not been observed in the radio emission of other pulsars except for the structure of giant pulses~\citep{2002A&A...396..171P}. On the other hand, the radio emission of our pulsar completely lacks any traces of a "normal" microstructure with time scales of about a millisecond!
The absence of such a structure in radio pulses is probably due to
the projection of the line of sight onto the magnetic meridian, so that each pulse reflects the development of a single process of secondary plasma generation,
which, as we see, turns out to be homogeneous. The idea arises that
the "normal" microstructure of other pulsars is due to the inhomogeneity of the structure of the secondary plasma in the transverse direction to the magnetic meridian.
Indeed, the pulsar B0809+54, in which the observer's line of sight passes along the very edge of the cone of open magnetic field lines
almost perpendicular to the magnetic meridians, a rich microstructure of millisecond and submillisecond scales was found (\citep{1982SvA....26..439P, 1981A&A....93...85B, 1981SvA....25..442S}).
With such an interpretation of the nature of the microstructure, the dependence of the time scale of the microstructure on the rotation period of the pulsar should be observed.
This dependence was noted in the work of Kramer et al.~\citep{2002MNRAS.334..523K} and was discussed in publications~\citep{2019MNRAS.483.4784M,Bartel2002}.

\section{Conclusion}\label{conc}

1. Analysis of our data has shown that components 1, 2, and 5, 6 of the average profile are formed by an ordinary radio emission mode (O mode) and form a single cone radiation of the pulsar. The splitting of this cone into two components occurs due to the competition of discharges generating secondary plasma over the polar cap of the pulsar. The idea of such a model for the formation of an average profile was proposed by Gil and Sendik~\citep{2000ApJ...541..351G},

2. The central components of the average profile (components 3 and 4) are formed by an extraordinary  mode (X-mode). A detailed justification for this view is given in  publication by Smith and co-authors~\citep{2013MNRAS.435.1984S} based on an analysis of the behavior of circular polarization. Our estimates of the height of the output of the central radiation (X-mode) and the cone radiation (O-mode) give values of 80~km and 370~km, respectively.

3. In the center of the middle profile, we found a new component with the narrowest half-width (1.5 - 2.5~ms). We believe that this component is formed as a result of the summation of radiation from a circular central zone with a diameter of about one degree. In this zone, the line of sight is close to the direction of the magnetic field lines within the final radiation angle with a relativistic factor $\gamma$ of about 200. Such a summation of radiation can explain the anomaly of the behavior of the positional angle of the plane of linear polarization, noted in the work~\citep{2013MNRAS.435.1984S}.

4. For pulsar B1237+25, there is an instability in the process of secondary plasma generation over the polar cap. Therefore, it is characterized by the phenomena of nulling, subpulse drift, and a change in the radiation mode.  The subpulse drift, however, is observed in the normal QN radiation mode only in the extreme components 1 and 6. The normal QN radiation mode is interrupted by nullings and transitions to the abnormal AB mode. The anomalous mode reflects another stable configuration of the activity zones. At the same time, the structure of the activity zones at the edge of the outer cone collapses (the carousel stops working), the distance between the inner and outer cones almost doubles from 8.3~ms to 16.3~ms (see Table 1), and the distance between the inner cone and the core is reduced and filled with radiation. There is a significant increase in the compact component 4, since, in our opinion, this component integrates radiation from the entire ring of the nearest activity zones. 

5. A microstructure with a submicrosecond time scale of $\tau_\mu\le0.5$~microseconds has been detected. This time scale corresponds well to the characteristic time of the development of a spark discharge in the polar cap. The characteristic discharge development time is $\tau_\mu{=}\Lambda h_p/c$, where $\Lambda$ is a multiplicity factor of approximately 20~\citep{1982SvA....26..443B}. For the value of $\tau_\mu\le0.5$~microseconds, the height of the vacuum gap is $h_p\le750$~cm.\\
The absence of a "normal" pulse microstructure with a millisecond scale can be explained by the fact that such a normal microstructure, studied in other pulsars, reflects the spatial inhomogeneity of the secondary plasma clouds in the direction perpendicular to the magnetic meridian.

6. Based on the steepness of the individual pulse's trailing edge (Fig.~\ref{fig:separate}), a limit was obtained on the value of the $\gamma$ factor of the relativistic secondary plasma $\gamma\ge 260$.

7. The relative number of nullings and various radiation modes at a frequency of 111~ MHz turns out to be the same as at a frequency of 327~ MHz according to the work~\citep{2013MNRAS.435.1984S}. This confirms the broadband nature of these phenomena.

\bibliographystyle{raa}
\bibliography{B1237} 

\end{document}